\documentclass[11pt]{iopart}
\usepackage[utf8]{inputenc} % allow utf-8 input
\usepackage[T1]{fontenc}    % use 8-bit T1 fonts
\usepackage{url}            % simple URL typesetting
\usepackage{booktabs}       % professional-quality tables
\usepackage{amsfonts}       % blackboard math symbols
\usepackage{nicefrac}       % compact symbols for 1/2, etc.
\usepackage{microtype}      % microtypography
\usepackage{xcolor}         % colors
\usepackage{graphicx}
\usepackage{hyperref}       % hyperlinks
\usepackage{multirow}
\usepackage{varwidth}
\usepackage{orcidlink}
\usepackage{subcaption}

\expandafter\let\csname equation*\endcsname\relax
\expandafter\let\csname endequation*\endcsname\relax
\usepackage{amsmath}

\newcommand{\geant}     {\texttt{Geant4}}
\newcommand{\onbb}      {\ensuremath{0\nu\beta\beta}}

\newcommand{\siggen}    {\texttt{siggen}}
\newcommand{\ssd}       {\texttt{SolidStateDetectors}}

\newcommand{\cpunet}    {\texttt{CPU-Net}}

\newcommand{\ren}       {Response Emulation Network}
\newcommand{\autherabv} {K. H. Bhimani }
\newcommand{\gitlink}   {\url{https://github.com/aobol/CPU-Net}}

\begin{document}

\title[]{CycleGAN-Driven Transfer Learning for Electronics Response Emulation in High-Purity Germanium Detectors}

\author{Kevin Bhimani$^{1,2}$\footnote{Corresponding Author}\footnote{Present Address: University of Delaware, 591 Collaboration Way, Newark, DE 19713, USA}, Julieta Gruszko$^{1,2}$, Morgan Clark$^{1,2}$\footnote{Present Address: Naval Research Laboratory, Optical Sciences Division, 4555 Overlook Ave SW, Washington, DC 20375, USA}, John Wilkerson$^{1,2,3}$, Aobo Li$^{4}$}
\address{$^1$ University of North Carolina at Chapel Hill, 120 E. Cameron Ave., Chapel Hill, NC 27599, USA}
\address{$^2$ Triangle Universities Nuclear Laboratory, 116 Science Dr., Durham, NC 27708, USA}
\address{$^3$ Oak Ridge National Laboratory, Oak Ridge, TN 37830, USA}
\address{$^4$ University of California, San Diego, 9500 Gilman Dr., La Jolla, CA 92093, USA}

\ead{kevinhbhimani@gmail.com}
% \vspace{10pt}
% \begin{indented}
% \item[]August 2017
% \end{indented}

\begin{abstract}
High-Purity Germanium (HPGe) detectors are a key technology for rare-event searches such as neutrinoless double-beta decay ({\onbb}) and dark matter experiments. Pulse shapes from these detectors vary with interaction topology and thus encode information critical for event classification. Pulse shape simulations (PSS) are essential for modeling analysis cuts that distinguish signal events from backgrounds and for generating reliable simulations of energy spectra. Traditional PSS methods rely on a series of first-principles corrections to replicate the effect of readout electronics, requiring challenging fits over large parameter spaces and often failing to accurately model the data. We present a neural network architecture, the Cyclic Positional U-Net ({\cpunet})\footnote{\gitlink}, that performs translations of simulated pulses so that they closely resemble measured detector signals. Using a Cycle Generative Adversarial Network (CycleGAN) framework, this {\ren} (REN) learns a data-driven mapping between simulated and measured pulses with high fidelity, without requiring a predetermined response model. We use data from a High-Purity Germanium (HPGe) detector with an inverted-coaxial point contact (ICPC) geometry to show that {\cpunet} effectively captures and reproduces critical pulse shape features, allowing more realistic simulations without detector-specific tuning. {\cpunet} achieves up to a factor-of-four improvement in distribution-level agreement for pulse shape parameter reconstruction, while preserving the topology-dependent information required for pulse-shape discrimination.
\end{abstract}

\section{Introduction}\label{sec:intro}
\begin{figure}[!htbp]
\centering
  \raisebox{2ex}{\includegraphics[height=0.24\textheight]{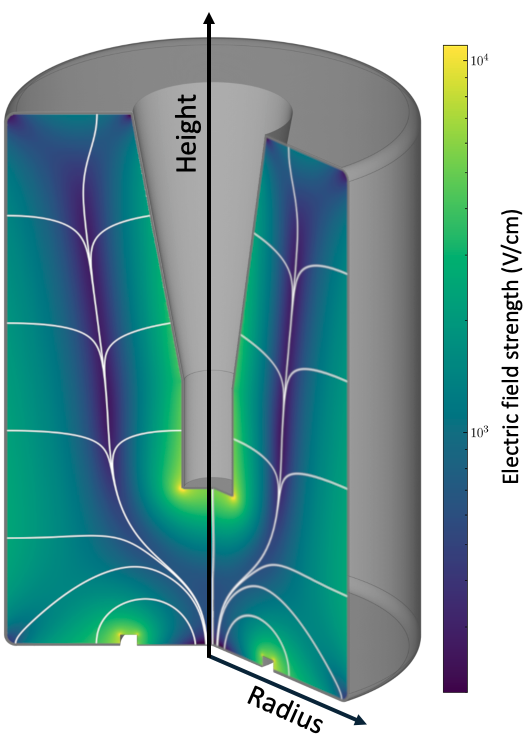}}
    \includegraphics[height=0.24\textheight,trim={0pc 0pc 1pc 0pc},clip]{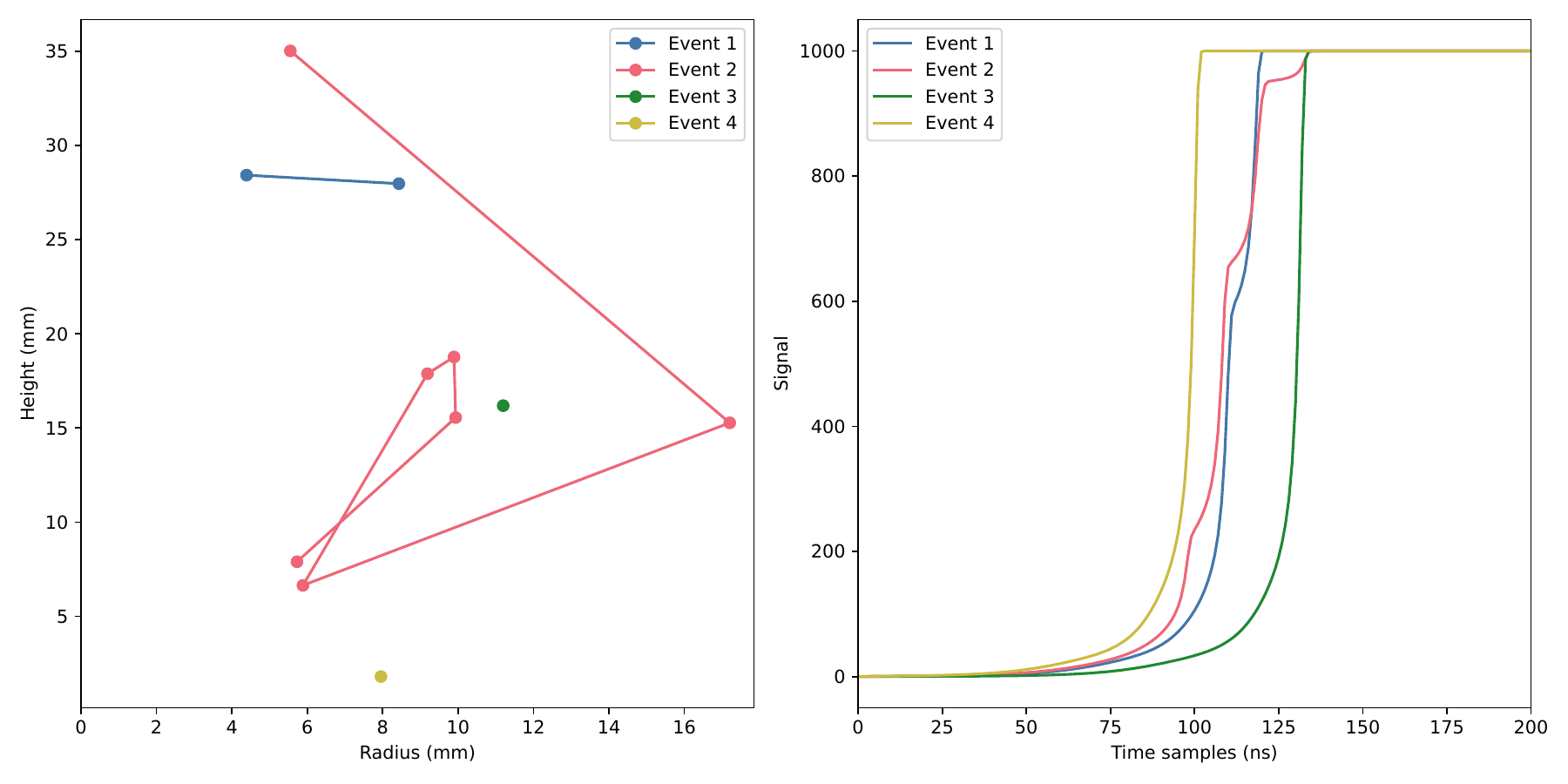}
    \caption{\textit{Left:} A cross section of the ICPC detector used, with the color bar depicting electric field magnitude and the white lines showing possible charge drift paths. The detector has $40.5$ mm radius and $96.0$ mm height. Image adapted with permission from the LEGEND Collaboration.
    \textit{Center:} Four events in the detector, simulated using {\geant}. \textit{Right:} Pulse shape simulations, generated using {\siggen} software, corresponding to the four events shown. Pulses are aligned by their start time and normalized to the total event energy. 
    }
   \label{fig:eng_dep_sim}
\end{figure}

High-Purity Germanium (HPGe) detectors play a crucial role in numerous low-background experimental searches for neutrinoless double-beta decay and dark matter~\cite{LEGEND_2025, MJD, GERDA, SuperCDMS, COGENT}. In these experiments, the HPGe detectors used often have a point contact geometry, which features a small readout electrode, called a point contact, at one end of the crystal insulated from the outer surface which is held at high voltage. This layout lowers the detector capacitance, leads to relatively long charge drift times (of $\mathcal{O}(\mu s)$) and yields a very steep gradient in the weighting-field near the point contact. In combination, these features produce pulse shapes that are highly sensitive to the energy deposition position, allowing for multi-site and surface event discrimination based on detector pulse shapes. These techniques are broadly termed pulse-shape discrimination (PSD). A variant of the point contact design is the inverted-coaxial point contact (ICPC) detector, which combines the low capacitance and excellent energy resolution of point-contact detectors with a larger active volume typical of coaxial designs. A cutaway view of an ICPC detector is shown in the left panel of figure \ref{fig:eng_dep_sim}.

The microphysics in the bulk of HPGe detectors is very well understood, allowing for accurate simulation of pulses. Charges, freed into the valence band by an energy deposition, drift under the electric field to the readout electrodes, inducing the measured signal. Pulse shape simulation (PSS) pipelines using publicly available tools such as \texttt{fieldgen} and {\siggen} \cite{icpc_siggen} or {\ssd} \cite{ssd_paper} software packages start from an electric field simulation that takes into account the geometry and impurity concentration of each detector. Each energy deposition taken from a particle interaction simulation tool such as {\geant} \cite{geant4one,geant4two, geant4three} is used to simulate the pulse induced by charge drift in the electric field and collection from the deposition site. The pulses are scaled according to the energies deposited at each site and summed together to create the total event pulse.  

The shape of the detector pulses is affected by many parameters such as particle energy, interaction position, and ionization track density. The relationship between the location of energy deposition and the resulting detector pulse is illustrated in figure~\ref{fig:eng_dep_sim}. The center panel depicts the hit locations of the particles for four events. In Event $1$ and $2$, the energy depositions occur at multiple spatially separated sites, resulting in multi-site events, typical of a gamma interaction. Event $3$ and Event $4$ are effectively single-site, as the energy depositions occur very close to each other and thus produce events in which the multiple depositions do not have a measurable effect on the resulting pulse, typical of an electron or charged particle interaction. Events $3$ and $4$ have different drift times because they are deposited at different distances from the point contact.

PSS is critical to designing high-performing detectors, understanding event topology, and distinguishing between signal and background events.~\cite{ MERTENS201981, Comellato_2020ljj}. PSS aims to generate pulses based on these parameters that are indistinguishable from detector pulses as measured in data. This allows us to produce ground-truth datasets where a mapping is established between each simulated pulse and its corresponding incident particle(s). Suppose we apply a pulse shape discrimination cut that is commonly used to identify and reject background events in {\onbb} experiments to this dataset, (see Ref.~\cite{AvsE}). In that case, it will be immediately apparent which event topologies lead to the signal sacrifice or background contamination of this cut. This information can be used for evaluation and further improvement of cut efficiency. Pulse shape simulation is also critical for designing new cuts and new HPGe detector geometries, as well as training machine learning models to replace or supplement traditional discrimination methods.

In point-contact-style detectors, the signal is read out at a small point-like ground electrode, with the resulting electric field leading to sharply rising detector pulses. This readout electrode is instrumented with an amplification and digitization chain, so a frequency-dependent response function must be added to the simulations to reproduce the true digitized detector signal. This electronics transfer function models the effect of the readout chain as a series of linear transformations on the output signal. The electronics response is typically derived from the circuit response to a step function input. However, the experimental conditions commonly used in low-background and cryogenic experiments, such as long cable runs and extended amplification chains, make it difficult to generate step-like pulses. This poses significant challenges in measuring the electronics response directly.

Although simulations of the readout electronics or approaches that fit the parameters of the electronics transfer function using data can account for some of these effects, they are inherently limited by assumptions. These assumptions often fail to capture the complexities of real-world behavior such as detector-specific anomalies, nonlinearities in the amplification chain, and time-varying and channel-specific effects from experimental conditions. Furthermore, inaccuracies in simulation assumptions, such as incorrect drift velocity assumptions in the bulk of the detector or oversimplified charge cloud dynamics, can exacerbate discrepancies between simulated and real detector pulses. The result is a persistent mismatch between the simulated and measured pulses, necessitating data-driven corrections. 

Current experiments avoid detailed electronics modeling by calculating reconstruction features directly from Monte Carlo particle-interaction simulations, relying on heuristic methods to meet most of their simulation needs \cite{legend_pcdr}. These methods can be tuned to match experimental data in a particular region of interest, but generally do not reproduce pulse shape behavior correctly for all spectral features. They also fall short in their ability to account for detector-by-detector variations and operational changes in experimental conditions, which can evolve over time. 

Transfer learning can help overcome the need for a first-principles electronics model and parameter retuning by directly learning the translations between each simulated pulse and its corresponding detector pulse. In this work, we present a novel transfer learning framework called Cyclic Positional U-Net~({\cpunet}) for improving pulse shape simulation. {\cpunet} translates simulated pulses into detector data-like pulses so that they closely resemble real detector pulses. We demonstrate that {\cpunet} improves the reproduction of distributions of key pulse shape parameters, such as current amplitude, rise times, and tail slopes, without including these distributions or the parameter values as targets of the network training.

\section{Cyclic Positional U-Net}\label{sec:cpu_net}
\subsection{Mathematical Formalism}
The task of learning the transfer function of electronics from data and adding its effects to the simulation can be redefined as a transfer learning problem. In this approach, the source domain $\mathcal{D}_{Source}$ represents the simulated pulse set. It includes features of the simulated pulses, such as energy, maximal current amplitude, tail slope, etc. These features form a feature space ($\mathcal{Y}$) and follow a probability distribution $(P(Y))$. The source task refers to obtaining reconstruction feature $(Y)$ in the source domain from the set of input simulated pulses, represented by $\mathcal{X}$. This is captured by the conditional distribution $P(X|Y)$, which describes how each parameter set $(Y)$ is assigned to a specific simulated pulse $(X)$. For instance, if we know the energy and drift time of a pulse, the source task models how these features produce a simulated pulse. Mathematically, we write the source domain as follows:

\begin{equation}
    \mathcal{D}_{Source}=\{\mathcal{Y},P(Y)\}\qquad 
    \label{eqn:source_domain}
\end{equation}
Then the reconstruction features are written as:
 \begin{equation}
     Y=\{\mathrm{E},I_{\mathrm{max}},c_{\mathrm{tail}},...\}\in \mathcal{Y}
 \end{equation}
The source task is written as:
\begin{equation}
    \mathcal{T}_{Source}=\{\mathcal{X},P(X|Y)\} \qquad X\in \mathcal{X}
    \label{eqn:source_task}
\end{equation}

Similarly, target domain $\mathcal{D}_{Target}$ represents the detector data pulses. It contains the feature space $\mathcal{Y'}$, whose elements follow the probability distribution $P(Y')$. $Y'$ are the reconstruction features of the detector pulses $(X')$.

\begin{equation}
\mathcal{D}_{Target}=\{\mathcal{Y}',P(Y')\}
\end{equation}
\begin{equation}
\mathcal{T}_{Target}=\{\mathcal{X}',P(X'|Y')\}
\end{equation}

The traditional pulse shape simulation approach attempts to approximate $P(X'|Y')$ by introducing nuisance parameters into $P(X|Y)$ and fitting $\mathcal{T}_{Target}$ to obtain their values. These parameters are applied to a given $Y$ in $\mathcal{D}_{Source}$ so that the set of simulated pulses $(\mathcal{X})$ matches the set of data pulses $\mathcal{X}'$. For example, one can fit the tail slope of the detector pulses to determine the decay constant and add a pole with that time decay to the simulation to replicate the data. This method requires a complicated collection of modeling and characterization data, along with computationally expensive fitting procedures in a highly degenerate high-dimensional parameter space \cite{Ben_Thesis,Sam_Thesis}. 

Instead, we can avoid the direct modification of $P(X|Y)$ by introducing a simulation translator. We call it the {\ren} (REN) represented by $\Lambda$:
\begin{equation}
\Lambda = \{\hat{\mathcal{X}}, P(\hat{X}\mid X)\}\qquad \hat{X}\in \hat{\mathcal{X}}
\label{eqn:REN}
\end{equation}
REN accepts an input pulse $X$ and translates it to an output pulse $\hat{X}$. This transformation is learned from a large sample of data pulses. The collection of transformed output $\hat{\mathcal{X}}$ should be very similar to $\mathcal{X}'$ after training, so that by combining the REN and $\mathcal{T}_{Source}$, we can replicate $\mathcal{T}_{Target}$:
\begin{equation}
    \mathcal{T}_{Target}=\Lambda \mathcal{T}_{Source}
    \label{eqn:REN_task}
\end{equation}
Similarly, we define a data translator called the Inverse {\ren} (IREN) represented by $\bar{\Lambda}$.
\begin{equation}
    \mathcal{T}_{Source}= \bar{\Lambda} \mathcal{T}_{Target}
    \label{eqn:IREN_task}
\end{equation}
which allows us to learn the features of $\mathcal{T}_{Target}$ without explicit programming. 

We intentionally do not impose an explicit loss on the target‐domain feature distribution $P(Y')$. Constraining the network solely at the waveform level ensures it remains agnostic to any particular set of pulse parameters. This allows newly devised pulse-shape based analysis techniques to be studied without bias. Moreover, the electronic response is a global effect that should correct features common to all pulses. By not fitting $P(Y')$ directly, the translator avoids compensating for event‑by‑event physics differences, allowing residual disagreements between simulations and data to serve as clean diagnostics of the systematic uncertainty associated with PSS methods. In short, the goal of CPU‑Net is to perform a translation of each simulated waveform rather than generating entirely new pulses.

A limitation of this approach is that electronics chain introduces stochastic effects and other effects that destroy information, e.g. electronic noise and limited bandwidth at high frequency. Therefore, an invertible mapping between simulation and data does not exist. Regardless of this limitation, this method provides a very good approximation for practical application, as the pulse shape parameters of interest are designed to be as insensitive to noise and high-frequency effects as possible. 

\subsection{Network Architecture}

\subsubsection{Positional U-Net}
\begin{figure}[!htbp]
    \centering
    \includegraphics[width=0.59\linewidth]{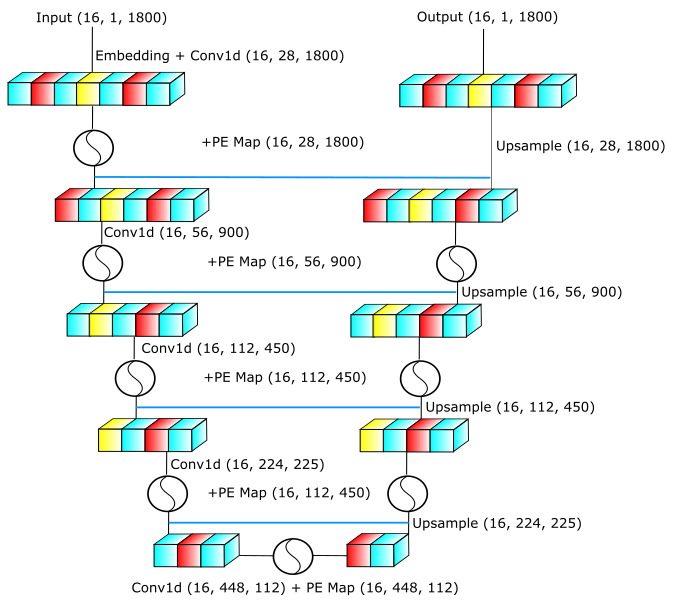}
    \includegraphics[trim={5cm 6cm 5cm 6cm},clip,width=0.40\linewidth]{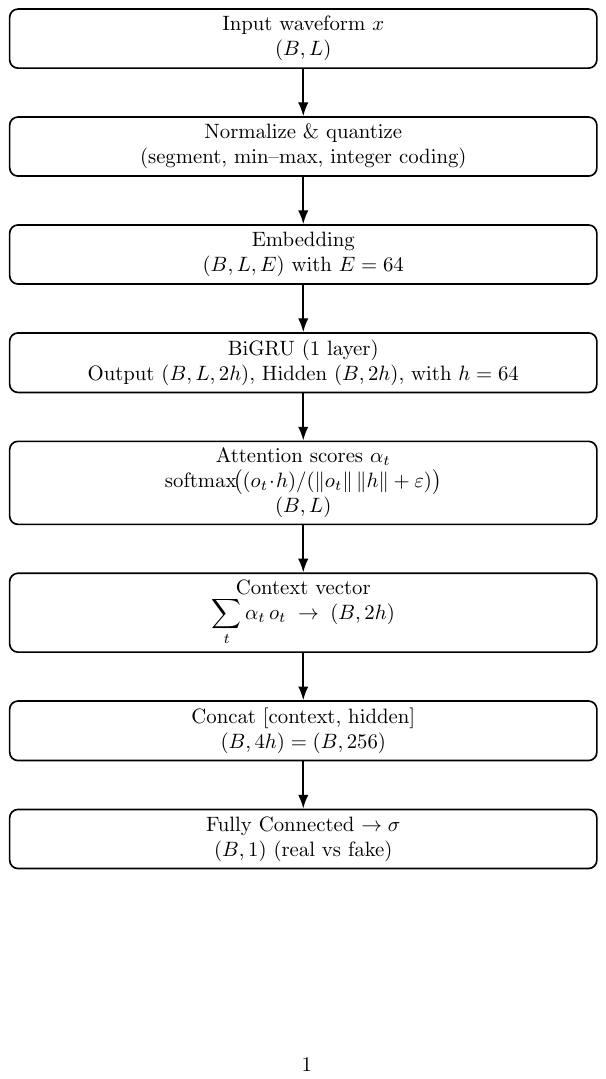}
    \caption{\textit{Left:} Layer-wise breakdown of the Positional U-Net. The blue lines represent the skip connections between contracting and expanding paths of the U-Net. Positional encoding layers are also shown for all levels. \textit{Right:} Flow of pulses in the RNN+Attention discriminator. The embedded waveforms pass through a bidirectional GRU and attention layer to form context-aware features. The output is concatenated with hidden states from the GRU. This is fed to a fully connected network which produces a classification decision using a sigmoid function.}
    \label{fig:network_schematic}
\end{figure}

Our method relies on a U-Net \cite{UNet}, a convolutional neural network initially developed for biomedical image segmentation, as a baseline model when designing the REN. A U-Net contains contracting (encoding) and expansive (decoding) channels. The contracting path contains $n$ convolutional layers to encode a pulse into a characteristic vector to capture contextual information and reduce the spatial resolution of the input. The expanding path contains $n$ upsample layers to decode the feature vector back to an output with the same length. The network structure also allows information to flow at different levels to the decoding part to provide maximum reconstruction efficiency. The U-Net structure is depicted in figure~\ref{fig:network_schematic}, where the tensor shape is also denoted at each stage. The Conv1d module in the U-Net is a series of layers; within each layer, the kernel size is an important hyperparameter to control the receptive field of convolutional layers, and padding is added to guarantee the same input and output shapes. Max pooling is used in all Conv1d modules except for the first layer to reduce the feature map size. This increases the effective receptive field by focusing on the most essential features in the signal.

%trim= left bottom right top

\begin{figure}[!htbp]
    \centering
    \includegraphics[height=0.7\textheight]{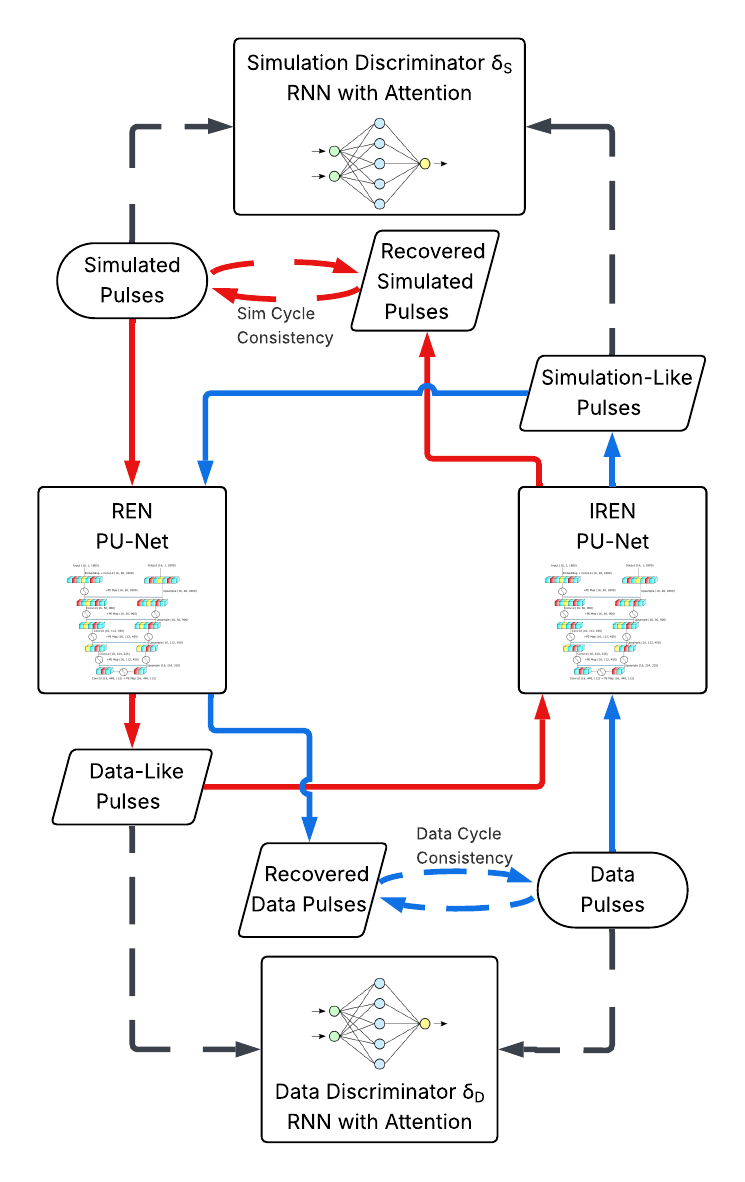}
    \caption{A flowchart representing the cycle-consistent adversarial training of the Cycle-GAN network. The solid lines indicate the flow of training instances through the PU-Net {\ren}~(REN) and Inverse {\ren}~(IREN) generators, and dotted lines represent how each stage is used to produce a subset of the losses used to train the network. The full list of the losses used is given in Table~\ref{tab:loss_summary}. The red lines represent the forward-cycle, which begins with simulated pulses and transfers them to data-like pulses using the REN. The Data Discriminator $\delta_D$ differentiates data-like pulses from true data and is used for adversarial training of the REN. Completing the cycle through the IREN to produce recovered simulation pulses allows the use of an a weighted mean absolute error (L1) loss to ensure cycle consistency. The reverse cycle, shown by the blue lines, is trained simultaneously with the same strategy, beginning with data pulses.}
    \label{fig:cycle-gan}
\end{figure}
During network design, we observed that the conventional U-Net does not reproduce the tail of the pulse due to the lack of positional information in intermediate layer outputs. Therefore, we developed a Positional U-Net~(PU-Net) model with layer-wise positional encoding maps $\mathcal{M}_{\mathrm{position}}$ inspired by the Transformer model~\cite{vaswani2017attention}. $\mathcal{M}_{\mathrm{position}}$ contains sine and cosine functions with different frequencies. Since each U-Net layer outputs a tensor with a different shape, $\mathcal{M}_{\mathrm{position}}$ must be generated separately for each layer and added to the layer output, as shown in the left panel of figure~\ref{fig:network_schematic}.

Finally, we added a reparameterization trick to the bottom of PU-Net. The trick was developed by Kingma and Welling \cite{VAE} to sample a random space while preserving the gradient flow. Reference \cite{AAE} pointed out that this trick has the ability to increase the stochasticity of machine learning models. Based on our experiments, increased stochasticity helps the model learn the distributions of the reconstruction features better. We call this model Positional U-Net (PU-Net). Together, the PU-Net model has $7,213,781$ trainable parameters.

\subsubsection{RNN with Attention} \label{subapp:RNN}

A RNN is a natural discriminator for pulses, because it can intrinsically preserve the temporal ordering of a waveform. We chose an attention-coupled recurrent neural network (RNN)\cite{bahdanau2015neural} as the discriminator for {\cpunet}; its overall structure is shown in the right panel of figure~\ref{fig:network_schematic}. A single-layer bidirectional RNN is used as the discriminator model in {\cpunet}. We adopt the Gated Recurrent Unit (GRU)~\cite{cho2014learning} as the internal structure of the RNN. The raw input pulses are first embedded in the $m=128$ space. Embedding is done by initializing a lookup table that maps integer ADC values to dense vectors of fixed size. These vector representations are learned during training iterations to capture meaningful pulse features. We use an embedding trick that optimizes computation by directly retrieving the corresponding row from the embedding matrix instead of performing unnecessary matrix multiplications. 

The embedded pulse is fed into a GRU that processes it sequentially while maintaining hidden states to capture temporal dependencies. This yields a $64$-dimensional output $\vec{I}(t)$ at each intermediate step $t$ as well as a final output $\vec{F}$. We then use an attention mechanism \cite{bahdanau2015neural} which can boost the performance of RNN by allowing it to focus on different parts of the pulses, such as the rising edge, during different steps of the training. The attention mechanism contains an attention matrix $A$ of dimension ($64$,$64$), which is used to calculate the attention score between $\vec{F}$ and each $\vec{I}(t)$. This calculates how much each time step contributes to the final output:
\begin{equation}
    s(t) = \mathrm{Softmax}[\vec{I}(t) A \vec{F}]
\end{equation}
A context vector is produced by summing $\vec{I}(t)$ with the weight $s(t)$ at each $t$. Finally, the context vector and the final output vector are concatenated and fed into a fully connected layer that produces a single scalar output. This model has $130,817$ trainable parameters.

\subsection{CycleGAN}
Ideally, the REN would be trained with paired pulses of simulations and data. In reality, collecting such a paired dataset ranges from a major challenge to functionally impossible, depending on the desired population of events. Although we can easily obtain $Y$ from $X$ through simulation, obtaining $Y'$ from $X'$ is difficult without precise knowledge of $P(X'|Y')$. This means it is impossible to obtain a paired dataset between $\mathcal{D}_{Source}$ and $\mathcal{D}_{Target}$.
 
The Cycle Generative Adversarial Network (CycleGAN) framework \cite{CycleGAN, hoffman2018cycada} provides an unsupervised learning approach to train the REN using adversarial losses. We first construct two networks: a REN~$\Lambda$ and an Inverse REN $\bar{\Lambda}$, both with PU-Net structure. We then construct two discriminator networks $\delta_{S}$ and $\delta_{T}$ for the source and target pulses, respectively, each based on a RNN with attention. We call this model, which combines all the features, the Cyclic Positional U-Net ({\cpunet}). The overall structure of the CycleGAN training framework is depicted in figure~\ref{fig:cycle-gan}; a detailed description of the framework, losses, and training is given in section~\ref{ssec:network_training}. The trained {\cpunet} produces both a REN and an IREN, translating pulses between the source and target domains. The REN is the primary interest of this work; future work will study whether the IREN can be used to improve data analysis methods.

\section{Data Preparation \& Model Training}\label{chap:training}
\subsection{Data Selection}
\begin{figure}
    \centering
    \includegraphics[height=0.26\textheight,trim={1pc 0pc 0pc 0pc},clip]{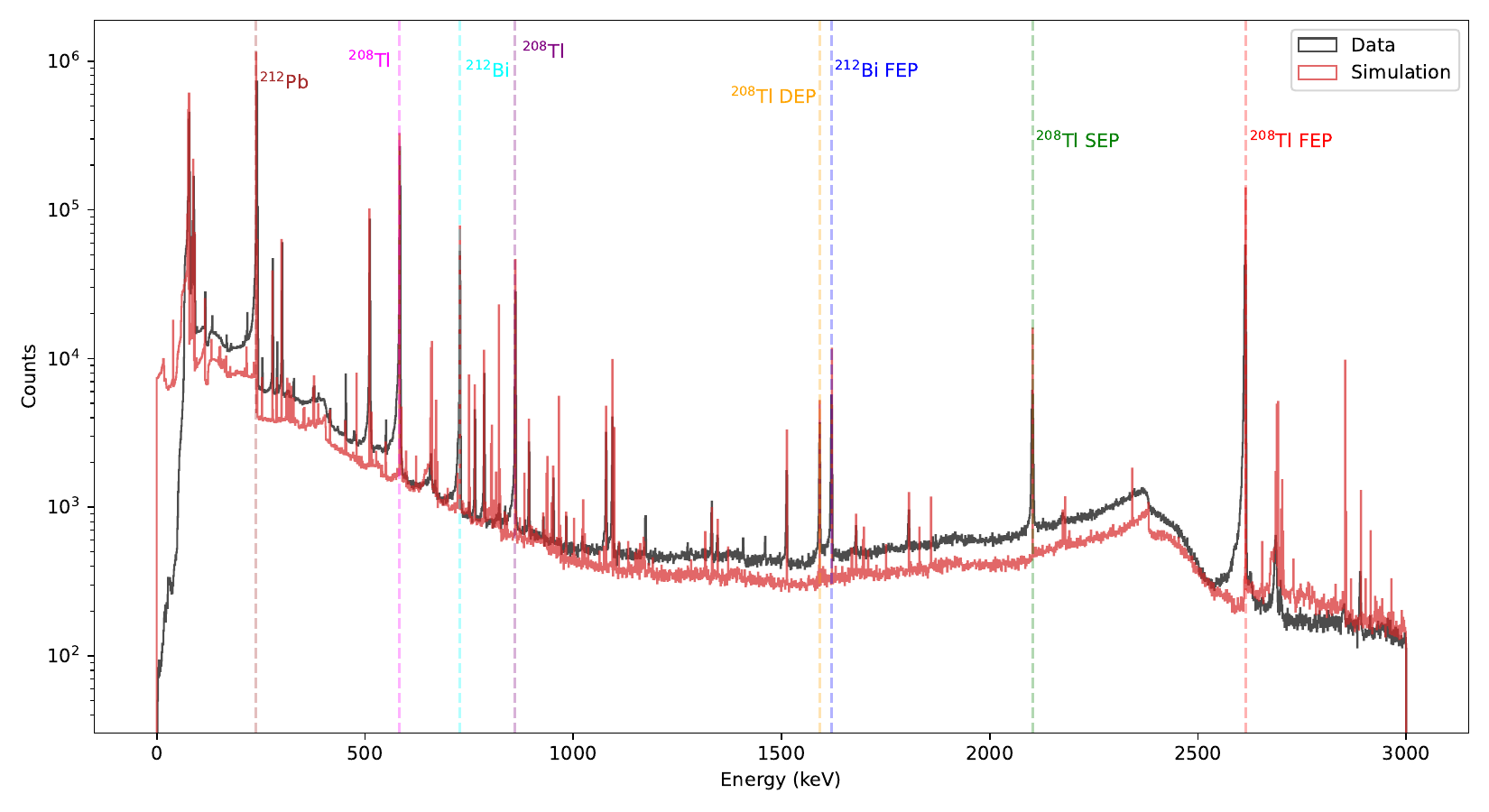}
    \includegraphics[height=0.26\textheight]{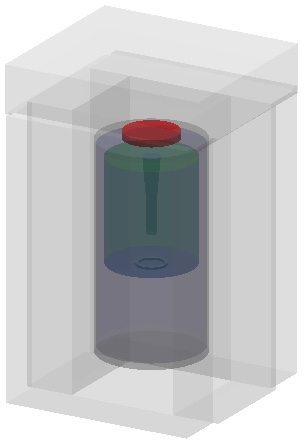}
    \caption{\textit{Left:} Calibrated energy spectrum from a $^{228}$Th source at ORNL compared to the Monte Carlo simulation results. Key peaks from $^{228}$Th source are labeled. The simulation does not include detector surface effects, nor any electronic‑noise or energy resolution effects, thus resulting in a mismatch. \textit{Right:} The simulation geometry of the ORNL characterization setup, showing the detector in green, the source in red, the aluminum cryostat and holder in dark grey, and the lead shielding in light grey.}
   \label{ch7_fig_eng_spec_comp}
\end{figure}

The LEGEND Collaboration characterized a subset of newly produced HPGe detectors at Oak Ridge National Laboratory (ORNL). For this study, we selected characterization data from a LEGEND Inverted-Coaxial Point-Contact (ICPC) detector, V06643A, manufactured by ORTEC \cite{ortec}. In the ORNL setup, the detector was mounted in a vendor cryostat, shielded by lead, and housed in a concrete alcove to reduce the external background. A Thorium-$228$ source was placed on top of the cryostat. The resulting detector signals were digitized using a FlashCam digitizer, and the \texttt{pygama} package \cite{pygama} and \texttt{dspeed} package \cite{dspeed} were used to convert the FlashCam output to LEGEND HDF5 files. The event energies were calibrated to convert the ADC units to keV. The energies were then corrected for charge trapping~\cite{mjd_pole_zero}. Figure \ref{ch7_fig_eng_spec_comp} shows the final energy spectrum obtained from the detector.

The peaks of interest for this study are the escape peaks. These are created from pair production events by a high-energy gamma ray. In this process, the incident photon generates an electron-positron pair. The electron creates a localized cascade and has its full energy deposited at a single site in the detector while the positron is slowed and annihilates, producing two additional gamma-ray photons. If all secondary gamma rays are fully absorbed within the detector, their combined energy yields an event at the energy of the initial gamma, contributing to the Full Energy Peak (FEP). The FEP spectrum includes pair-production events and events produced via other processes in the detector that may include only a single site, such as photoelectric absorption. Therefore, it includes a mixture of single-site and multi-site events.

If one of the annihilation gamma rays leaves the detector without depositing its energy, the spectrum shows a Single Escape Peak (SEP) event. This scenario results in a multi-site event because the signal is comprised of two distinct interaction sites within the detector: the pair production site, and the absorption site of one of the two gamma rays. If both gamma rays escape the detector without interacting, a Double Escape Peak (DEP) event is observed. This situation corresponds to a single-site event since only the pair-production site contributes to the signal. Tl-208 isotope is part of the decay chain of $^{228}$Th and produces an FEP at $2614.53$ keV. Given that the mass of the electron is $511$ keV/c$^2$, the SEP occurs at $2103.53$ keV and the DEP at $1592.53$ keV. 

The training losses are expectations over the empirical training samples and, therefore, depend on the choice of the training set. We intentionally train on the high-statistics Full Energy Peak (FEP), which contains a mixture of single-site and multi-site topologies. This choice enables the network to learn a translation that is common across topologies rather than fitting topology-specific physics. To test this generalization, we validate it on the SEP (multi-site–dominated) and the DEP (single-site–dominated) regions of the spectrum. These are not perfectly pure samples, as Compton scattering introduces an underlying continuum of background events, and bremsstrahlung creates a population of multi-site DEP events. However, they are dominated by single-site (in the DEP) and multi-site (in the SEP) topologies, offering appropriate populations to validate the performance of CPU‑Net.

\subsection{Simulations}
{\geant} was used to simulate particle interactions in the ORNL characterization setup. The geometry modeled in {\geant} is shown in figure \ref{ch7_fig_eng_spec_comp}. It includes a HPGe detector, a radioactive source, and an Aluminum PopTop cryostat holder surrounded by lead shielding. We simulated $100$ million $^{228}$Th decay events originating from the source and recorded their energy depositions within the HPGe detector. 

Given the Geant4 simulations of energy depositions within the detector, we used {\siggen} simulations to generate pulses for hits in the DEP, SEP, and FEP energies. The {\siggen} simulations treat every energy deposit as a single point charge and heuristically account for diffusion and self-repulsion by convolving the output signal with a Gaussian that mimics the effects of a charge cloud with 0.1\,mm full-width at half-maximum (FWHM). We disabled the preamplifier decay in the simulation. The simulation does not include detector surface effects, which can degrade event energies in regions within $1$-$2$ mm of the detector contact surfaces. Electronic noise and other detector-level effects affecting the energy resolution are also not included in the simulations. These cause spread in the measured spectrum due to statistical fluctuations in charge carrier statistics, recombination, and electronic noise. The simulated energy spectrum was obtained by energy-weighted summation of the pulses for energy depositions for a given event. Figure \ref{ch7_fig_eng_spec_comp} also shows the simulated spectrum.

\subsection{Post Processing}
Normalizing and aligning the pulses is crucial for training the network. The simulated pulses are already normalized between $0$ and $1$. The raw data pulses are normalized by dividing by the $80\%$ of the average of the last five samples. Pulses are then vertically shifted so that the average of the first $200$ samples is zero. This ensures that all pulses have their preamplifier decay tails and their baselines aligned, which can help the model learn the features better. We used the $99.9\%$ rise time to horizontally align the pulses. We found this method more effective for training the network than aligning by the start of the rising edge, as the pulse shape at the end of the rise time is far more consistent between events than the shape of the beginning of the charge drift, which depends on the position of the energy deposition. The simulated pulses are padded to ensure that there are $400$ samples on both sides of the $99.9\%$ rise time. The preamplifier decay time constant $\tau$ of the data pulses is calculated by the slope of a linear fit of the logarithm of the last $300$ pulse samples. Events with poor fit quality $\chi^2$ or anomalous values of  $\tau$ are used to identify pile-up pulses in the data and remove them from training.

\subsection{Network Training}\label{ssec:network_training}
The training and validation of {\cpunet} are conducted in PyTorch~\cite{pytorch}, following the approach shown schematically in figure~\ref{fig:cycle-gan}. We define two PU-based generators called REN and IREN. Then we define two RNN with attention discriminators called $\delta_{S}$ and $\delta_{T}$. Suppose we start with a simulated pulse $X$ and data pulse $X'$, with $\Lambda(X)$ the output of the REN generator and $\bar{\Lambda}(X')$ the output of the IREN generator. During training, a simulated pulse $X$ is first fed to $\Lambda$ to produce a translated pulse $\Lambda(X)$. The discriminator $\delta_{T}$ attempts to distinguish $\Lambda(X)$ from real data pulses, while $\Lambda(X)$ attempts to `fool' $\delta_{T}$. Then $\Lambda(X)$ is fed to $\bar{\Lambda}$ to translate back to the simulated pulse source domain by $\bar{\Lambda}(\Lambda(X))$. The $X\rightarrow{}\Lambda(X)\rightarrow{}\bar{\Lambda}(\Lambda(X))$ translation path is termed the forward cycle, denoted by the red arrows of figure~\ref{fig:cycle-gan}. The same process performed in the other direction, starting with the data pulse $X'\rightarrow{}\bar{\Lambda}(X')\rightarrow{}\Lambda(\bar{\Lambda}(X'))$. A second discriminator $\delta_{S}$ attempts to distinguish $\bar{\Lambda}(X')$ from $X$, while $\bar{\Lambda}$ attempts to `fool' $\delta_{S}$. This is called the backward cycle, denoted by the blue arrows of figure~\ref{fig:cycle-gan}. The forward and backward cycles are trained simultaneously.

\subsubsection{Loss Functions}
Since we do not have a labeled training sample of matched data and simulation pulses, we cannot use a direct comparison loss. Instead, we define three loss terms that are optimized during training to help the network learn the translations.

The first of these is the $L_{\mathrm{Identity}}$ loss:
\begin{equation}\label{eq:loss_ided}
    L^{\mathrm{Id}}_{\Lambda} = \|X' - \Lambda(X')\|_1
\end{equation}

Since the goal of $\Lambda$ is to produce data-like pulses in $\mathcal{X'}$ domain, feeding in $X'$ to $\Lambda$ should return $X'$ itself. An equivalent identity loss term corresponding to the simulated pulses (i.e. for pulses $X$ and the IREN $\bar{\Lambda}$) is also included.

The cycle-consistent loss
\begin{equation}\label{eq:loss_cyc}
    L^{\text{Cyc}}_{\text{Fwd}} = \|X - \bar\Lambda\bigl(\Lambda(X)\bigr)\|_1
\end{equation}
 ensures that the circular translation path $X\rightarrow{}\Lambda(X)\rightarrow{}\bar{\Lambda}(\Lambda(X))$ preserves the original pulse shape. The forward cycle loss shown here corresponds to the case in which a simulated pulse is first fed into REN to produce a data-like pulse; this pulse is then fed through IREN to produce a simulation-like pulse, which should look exactly identical to the input simulated pulse. A similar loss exists for the backward translation path $X'\rightarrow{}\bar{\Lambda}(X')\rightarrow{}\Lambda(\bar{\Lambda}(X'))$. 
 
 In each case, the complete cycle should return the original input pulse. The cycle loss calculates the loss associated with this complete cycle.  The cycle-consistent loss acts in place of a direct comparison loss to ensure that the underlying physical features that are encoded in a pulse (e.g. multi-site vs. single-site, or energy deposition location) are not destroyed by the translation networks. Since the electronics response effect is not perfectly invertible, this loss will have a natural `floor' associated with the information loss of the true electronics response, even in the case of perfect network performance.  

There are two cycle and two identity loss terms to represent the forward and backward cycles. In these loss terms, the pulse is compared with another pulse. For such a comparison, we define a specialized mean absolute error (L1 loss) that emphasizes different parts of the pulse by assigning them varying weights. It is designed to give more importance to the rising and falling edges of the pulse, which are critical for accurate pulse shape analysis. 

In addition to the cycle and identity losses, there are four adversarial losses, one associated with each of the two generators and two discriminators. The adversarial losses calculate the loss associated with the generator and discriminator `fooling' each other. Each discriminator takes in the pulse from the generator and outputs a single value between 0 and 1. A value close to 1 means that the discriminator classifies it as a real pulse in its domain, while a value close to zero means that the discriminator classifies it as a translated (fake) pulse. A perfectly accurate simulated pulse (source domain) discriminator would produce the output:
$$ \delta_S(X) = 1;\hspace{4 pt} \delta_S(\bar{\Lambda}(X')) = 0 $$
and a perfectly accurate data pulse (target domain) discriminator would produce the output:
$$ \delta_T(X') = 1; \hspace{4 pt} \delta_T(\Lambda(X)) = 0$$
The adversarial losses for the discriminators are designed to penalize deviation from this behavior (i.e., they reward correct classification). The adversarial losses for the generators are designed to instead reward successful ``fooling'' of the discriminators, so they penalize correct classification.
 
The adversarial losses are quantified using the binary cross-entropy loss. For example, for discriminator $\delta_T$ the adversarial loss is:
 \begin{equation}
\label{ch7_eq_loss_gan}
    L^{\text{adv}}_{\delta_T}
  = -\mathbb E_{X'}\bigl[\log\bigl(\delta_T(X')\bigr)\bigr]
    - \mathbb E_{X}\bigl[\log\bigl(1-\delta_T(\Lambda(X))\bigr)\bigr]
\end{equation}
$\mathbb E_{X'}$ denotes the expectation value (batch average) over the set of real data pulses $X'$, and $\mathbb E_{X}$ is the expectation over the set of real simulated pulses $X$. The first term  penalizes~$\delta_T$ when it assigns a value smaller than~1 to a real data pulse~$X'$, and the second penalizes it when it assigns a value larger than~0 to a translated (fake) pulse~$\Lambda(X)$. An equivalent adversarial loss is calculated for the source domain discriminator $\delta_S$.

Conversely, the adversarial loss for the REN generator $\Lambda$ is obtained by passing a translated simulation pulse $\Lambda(X)$ through the data discriminator $\delta_T(\Lambda(X))$, so that it is encouraged to produce a pulse that the discriminator classifies as data. The REN generator $\Lambda$ tries to fool $\delta_T$ into believing that its translated output $\Lambda(X)$ is real data. In this case, the binary cross-entropy loss reduces to:
\begin{equation}
  \label{eq:adv_gen}
  L^{\text{adv}}_{\Lambda}
  = -\mathbb E_{X}\bigl[\log \bigl(\delta_T(\Lambda(X))\bigr)\bigr]
\end{equation}
with an equivalent adversarial loss calculated for the IREN generator $\bar{\Lambda}$.

The losses and optimizers are summarized in table \ref{tab:loss_summary}. A total of eight losses are optimized during training: an adversarial, cycle-consistent, and identity loss for each of the generators, yielding six losses associated with the generators, and an adversarial loss for each discriminator, yielding two losses associated with the discriminators. All losses are computed per pulse and then averaged once per batch. AdamW \cite{adam_w_paper} optimizers are used for all losses. Compared to the traditional Adam optimizer, AdamW applies weight decay as a separate step during gradient descent optimization. This avoids interference with the learning rate schedule and helps stabilize the training process. The six losses for REN and IREN are optimized together in a weighted sum using the weight hyperparameters given in table \ref{tab:hyperparameters}, and the discriminators each have their own optimizers. 

\begin{table}[htb]
  \centering
  \renewcommand{\arraystretch}{1.2} % Adjust row height for readability
\setlength{\tabcolsep}{4pt} % Adjust column spacing
  % \begin{tabular}{@{}llp{3.1cm}p{4.4cm}l@{}}

  \begin{tabular}{|p{0.19\linewidth}|p{0.19\linewidth}|p{0.1\linewidth}|p{0.305\linewidth}|p{0.13\linewidth}|}
  \hline
  \textbf{Networks} & \textbf{Loss name} & \textbf{Symbol} &
  \textbf{Analytic form} & \textbf{Optimizer} \\ \hline
  \multirow{6}{*}{Generators $\Lambda,\;\bar\Lambda$}
      & Data Identity        & $L^{\text{Id}}_{\Lambda}$ 
                               & $\|X' - \Lambda(X')\|_1$          & \multirow{6}{*}{AdamW\textsubscript{G}}\\
      & Sim Identity         & $L^{\text{Id}}_{\bar\Lambda}$ 
                               & $\|X - \bar\Lambda(X)\|_1$        &  \\
      & Forward Cycle            & $L^{\text{Cyc}}_{\text{Fwd}}$      
                               & $\|X - \bar\Lambda(\Lambda(X))\|_1$ & \\
      & Backward Cycle            & $L^{\text{Cyc}}_{\text{Bwd}}$      
                               & $\|X' - \Lambda(\bar\Lambda(X'))\|_1$ & \\
      & Data Adversarial     & $L^{\text{adv}}_{\Lambda}$        
                               & $-\mathbb E_{X}\!\bigl[\log( \delta_T(\Lambda(X)))\big]$ & \\
      & Sim Adversarial       & $L^{\text{adv}}_{\bar\Lambda}$     
                               & $-\mathbb E_{X'}\!\bigl[\log( \delta_S(\bar\Lambda(X')))\bigr]$ & \\[6pt] \hline
  Discriminator $\delta_T$     & Data Adversarial  
                               & $L^{\text{adv}}_{\delta_T}$        
                               & $\!\begin{aligned}
                                    &-\mathbb E_{X'}[\log(\delta_T(X'))]\\
                                    &\hspace{8pt}-\mathbb E_{X }[\log(1-\delta_T(\Lambda(X)))]
                                   \end{aligned}$ & AdamW\textsubscript{$\delta_T$} \\[6pt]\hline
  Discriminator $\delta_S$     & Sim Adversarial  
                               & $L^{\text{adv}}_{\delta_S}$        
                               & $\!\begin{aligned}
                                    &-\mathbb E_{X }[\log\delta_S(X)]\\
                                    &\hspace{4pt}-\mathbb E_{X'}[\log(1-\delta_S(\bar\Lambda(X')))]
                                   \end{aligned}$ & AdamW\textsubscript{$\delta_S$} \\ \hline
  \end{tabular}
    \caption{Loss terms optimized during CPU-Net training. One optimizer updates both generators using the weighted sum of their six losses, while each discriminator has an independent optimizer.}
  \label{tab:loss_summary}
\end{table}

\subsubsection{Hyperparameter Tuning}
Achieving stability in CycleGAN training can be challenging. This is because the loss optimization process is complex, with many metrics being optimized simultaneously. Training must balance the learning progress of generators and discriminators while preventing gradients from exploding to infinity or imploding to zero. To improve training, we introduced hyperparameters at different levels of the model that are summarized in table \ref{tab:hyperparameters}.

\begin{table}%[htb!]
\centering
\renewcommand{\arraystretch}{1.5} % Adjust row height for readability
\setlength{\tabcolsep}{2pt} % Adjust column spacing
\begin{tabular}{|p{0.22\linewidth}|p{0.1\linewidth}|p{0.63\linewidth}|}
\hline
\textbf{Hyperparameter}       & \textbf{Value} & \textbf{Description} \\ \hline
baseline\_len        & 200            & Samples in the pulse baseline window used for the weighted L1 losses. \\ \hline
rising\_edge\_len    & 250            & Samples in the pulse rising edge window used for the weighted L1 losses. \\ \hline
tail\_len            & 350            & Samples in the pulse decay tail window used for the weighted L1 losses. \\ \hline
baseline\_weight     & 3.0            & Weight given to the baseline of the pulse in L1 loss terms. \\ \hline
ris\_edge\_weight    & 10.0           & Weight given to the rising edge of the pulse in L1 loss terms. \\ \hline
tail\_weight         & 7.0            & Weight given to RC decay tail of the pulse in L1 loss terms. \\ \hline
batch\_size          & 32             & Pulses used in one training iteration to update model parameters. \\ 
\hline
iters                & 7000           & Number of iterations used for training. \\ \hline
decay                & 1000           & Iteration at which learning rate starts to decay. \\ \hline
lrate\_gen           & $1 \times 10^{-3}$ & Initial learning rate for the generator networks. \\ \hline
lrate\_disc          & $1 \times 10^{-3}$ & Initial learning rate for the discriminator networks. \\ \hline
cyc\_loss\_weight    & 20             & Weight of the cycle consistency loss terms in the overall generator loss.  \\ \hline
iden\_loss\_weight   & 5              & Weight of the identity loss terms in the overall generator loss. \\ \hline
gan\_loss\_weight    & 9              & Weight of the adversarial loss terms in the overall generator loss. \\ \hline
max\_grad\_norm      & 100            & Maximum gradient norm for gradient clipping. \\ \hline
w\_decay             & $1 \times 10^{-4}$ & Weight decay in the optimizers. \\ \hline
n\_disc\_iters       & 30             & Iterations after which the discriminators are updated. \\ \hline
\end{tabular}
\caption{Hyperparameters used for CPU-Net training. Following an initial hyperparameter space search using Bayesian optimization, these values were obtained through manual tuning to achieve stable training and good performance.}
\label{tab:hyperparameters}
\end{table}

We found that the discriminator typically overpowers the generator, since the generator has a more complex task of generating pulses while maintaining cycle and identity consistency and also fooling the discriminator. To balance this, we updated the weights of the generators more frequently than the discriminators. We introduced a hyperparameter for the number of intervals after which the discriminator is updated. This allowed the generator enough steps to adapt to changes in the discriminator without destabilizing the adversarial process.

Weighting the adversarial, cycle-consistency, and identity losses allows us to fine-tune the learning of the generators. At some point during the training, the generator will have learned just enough to fool the discriminator, and further learning might be hampered. The loss weights are used to push the generator to keep improving its translations: maintaining cycle and identity consistency—even after it has been successful in overpowering the discriminator. These hyperparameters were carefully tuned to achieve a balance, as values that are too high will cause the gradient to explode; too low will cause the model not to learn much. 

We also introduced a learning rate decay for the optimizers. In the beginning, the learning rate was intentionally kept high for the model to explore the entire parameter space. Then, the learning rate decays linearly to help it converge in the right direction. This was particularly important for the generator optimizer, as it is optimizing six losses, and ensures that all the space is explored.

To prevent overfitting during training, weight decay is applied to the optimizers to penalize large weights and ensure generalization among all parameters. Gradient clipping is applied, limiting the norm of gradients during training and preventing exploding gradients, particularly in the layers of the U-Net. The threshold for clipping the magnitude of gradients is kept high since the loss is multiplied by different loss weights multiple times, which can increase the magnitude.

Together, these parameters enable {\cpunet} to obtain the right balance during training. To optimize the hyperparameters, we conducted an initial search with Bayesian optimization \cite{balandat2020botorch} and then hand-tuned the hyperparameters to reach stable convergence. The hyperparameter optimization was performed with a fixed initial seed. A single training takes about 1 GPU hour on an NVIDIA A100 GPU. The trained {\cpunet} generates both a REN and an inverse REN, enabling bidirectional translation between the simulation and data domains. The REN is the primary focus of this work.

\section{Results}\label{sec:result}
Since labeled pairs of simulation and data pulses do not exist, the performance of this approach can only be quantitatively evaluated using information from distribution-level parameters. Individual pulse correspondences are used for a qualitative evaluation, however. We focused our quantitative studies of the results on distributions of pulse shape parameters associated with common analysis techniques used in low-background experiments.

In order to have accurate simulations, the REN needs to reproduce the correct ensemble distribution of the parameters used in PSD. Since dark matter signatures or {\onbb} decays would appear as single-site events, experiments often seek to distinguish between single-site and multi-site events using PSD techniques. We trained {\cpunet} on $110,000$ FEP pulses. With a batch size of $32$ and $7000$ training iterations, this enables the model to train on every pulse at least twice. We then used $12,000$ SEP events and $3,000$ DEP events for validation. In this section, we present the results of REN's ability to reproduce detector pulses by comparing the resulting distributions for some critical parameters used in PSD.

\subsection{Training Progression}
\begin{figure}
    \centering
    \includegraphics[width=\linewidth,trim={0.5cm 0pc 0.5cm 0pc},clip]{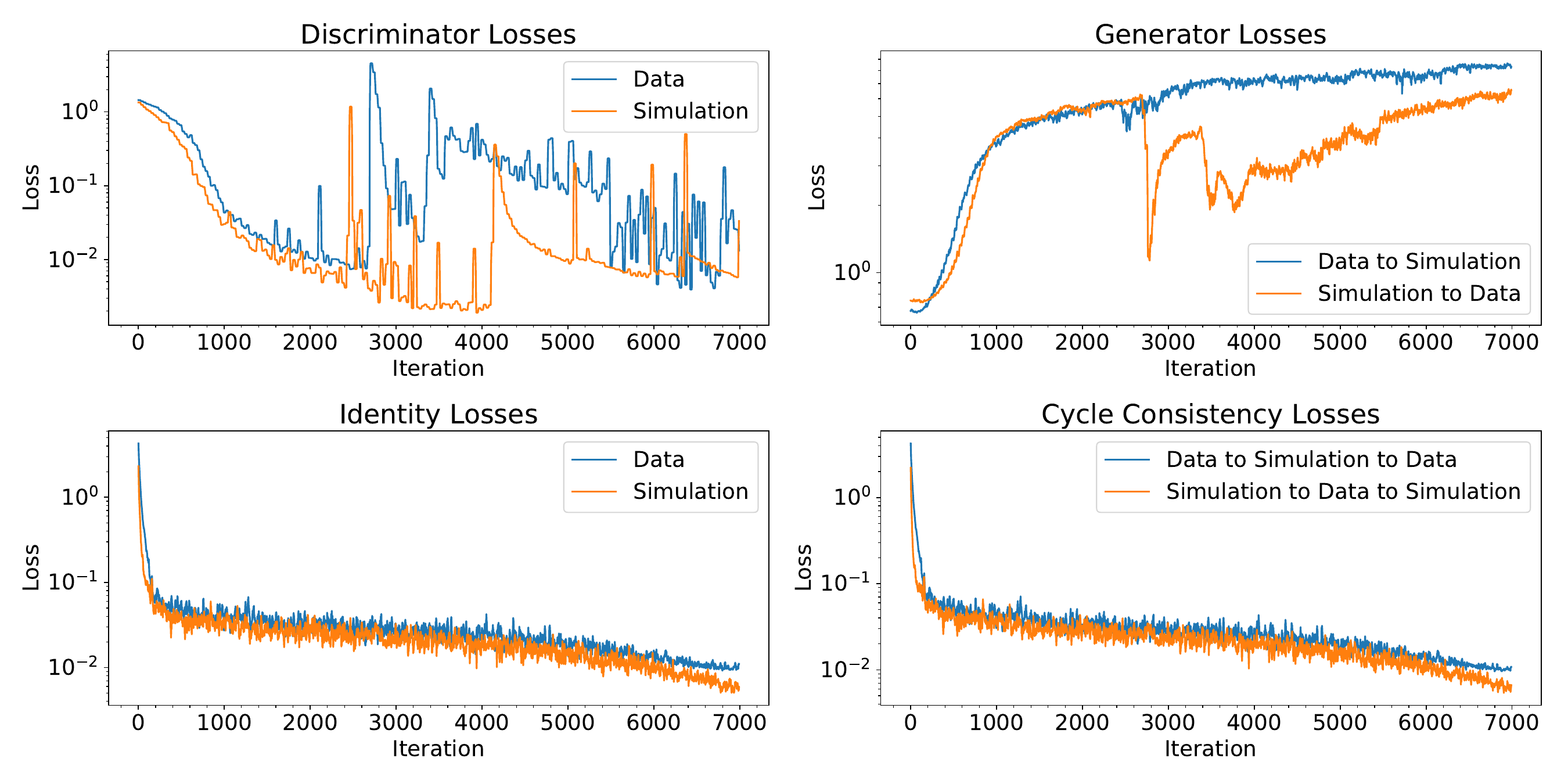}
    \caption{Training losses for {\cpunet}. Curves are smoothed using a moving average of 10 iterations for clarity. The identity and cycle losses rapidly converge, indicating that the translators learn to preserve the cycle and identity within the first few hundred iterations. Following an initial training phase in which the discriminator dominates, the generator and discriminator losses oscillate. This adversarial dynamic helps both models to learn and improve each other.} 
   \label{fig:training_loss}
\end{figure}

Figure \ref{fig:training_loss} shows the evolution of all eight losses (two for adversarial generators, two for adversarial discriminators, two for cycle consistencies, and two for identities), plotted with a 10-iteration moving average. The cycle and identity losses drop rapidly in the first few hundred iterations and remain small thereafter. The adversarial losses exhibit the expected GAN oscillations: as the discriminators momentarily improve, their losses decrease while the generators’ increase. Then, the trend reverses as the generators adapt. Given that the linear learning-rate decay begins at iteration 1000, and that the discriminators are updated every 30 generator steps, the system enters a regime in which the adversarial losses fluctuate around stable means. The jumps in the losses beyond 3000 iterations, therefore, indicate the alternating advantage typical of adversarial training, not divergence. 
 
\subsection{Pulse Translation}

\begin{figure}[!htbp]
        \centering
        \begin{subfigure} {0.99\textwidth}
            \centering
    \includegraphics[width=0.99\linewidth]{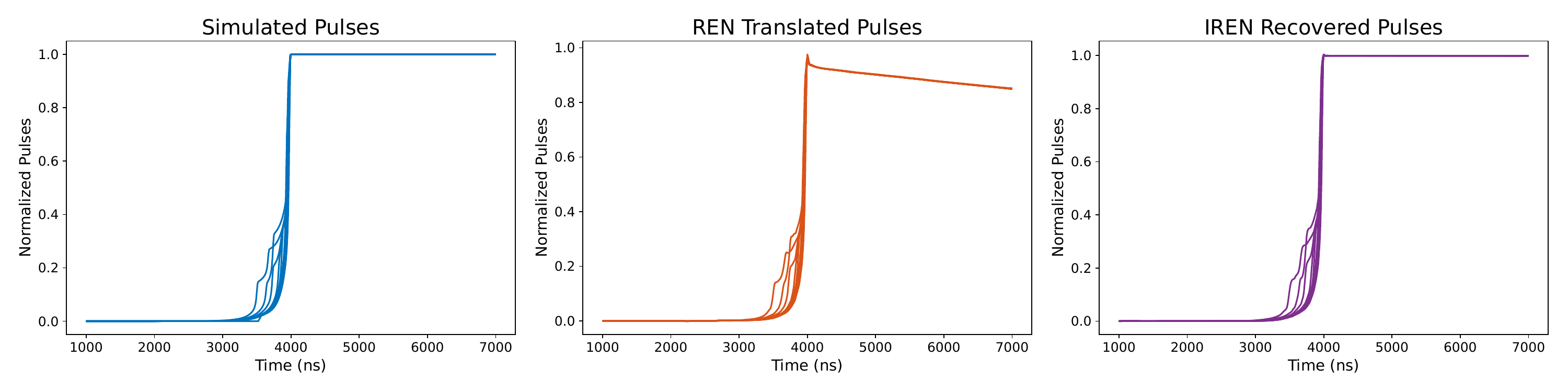}
    \caption{Pulse translation in the forward direction: starting with simulated pulses, the REN first translates them into detector-like pulses, and then IREN translates them back to the simulation-like pulses.}
    \label{fig:cycle_bab}
    
        \end{subfigure}
        
        \begin{subfigure} {0.99\textwidth}
        \centering
        \includegraphics[width=0.99\linewidth]{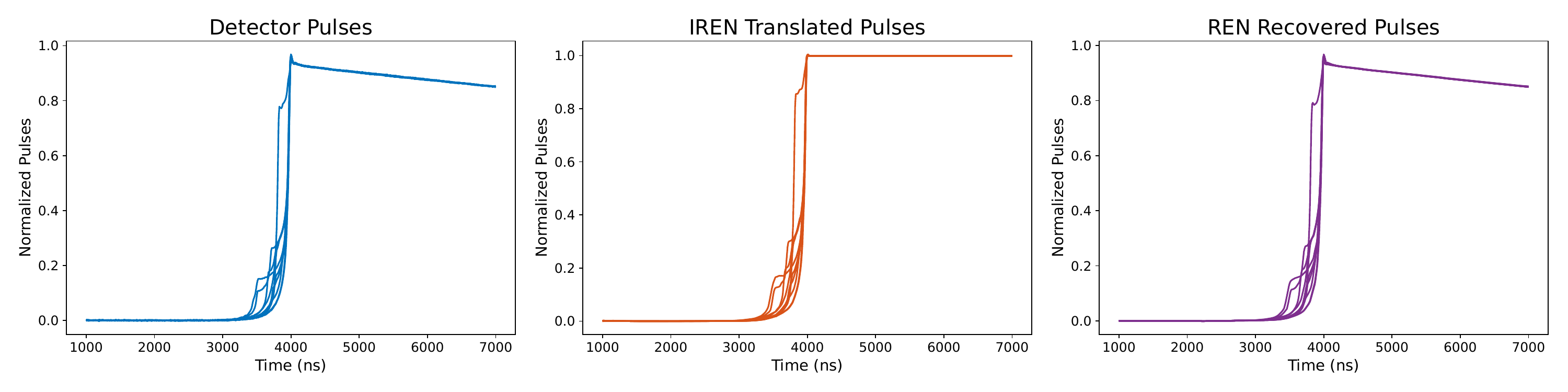}
    \caption{Pulse translation in the backward direction: starting with detector pulses, the IREN translates them into simulation-like pulses, and then REN translates them back to detector-like pulses.}
    \label{fig:cycle_aba}
        \end{subfigure}
        \caption{Examples of pulse translation through the {\cpunet} cycles for ten single-escape peak pulses.}
        \label{fig:pulse_examples}
        
    \end{figure}

Figures \ref{fig:cycle_bab} and \ref{fig:cycle_aba} show the SEP pulses as they progress through the forward and backward cycles in {\cpunet}, respectively. These SEP pulses feature a variety of event topologies, and the network effectively translates the pulses in forward and reverse translations while retaining this information. In the {\siggen} simulations, the RC discharge effect is not modeled, resulting in a flat tail slope of zero. REN learns to transform this flat tail into an exponentially decaying one while including more subtle effects like overshoot and a faster decay early in the tail, matching the observed behavior in real detector data, and IREN learns to transform pulse tails to a zero slope, matching the simulations. 

\begin{figure}[!htbp]
\centering
\includegraphics[width=0.8\linewidth,trim={0.0cm 0pc 0cm 0pc},clip]{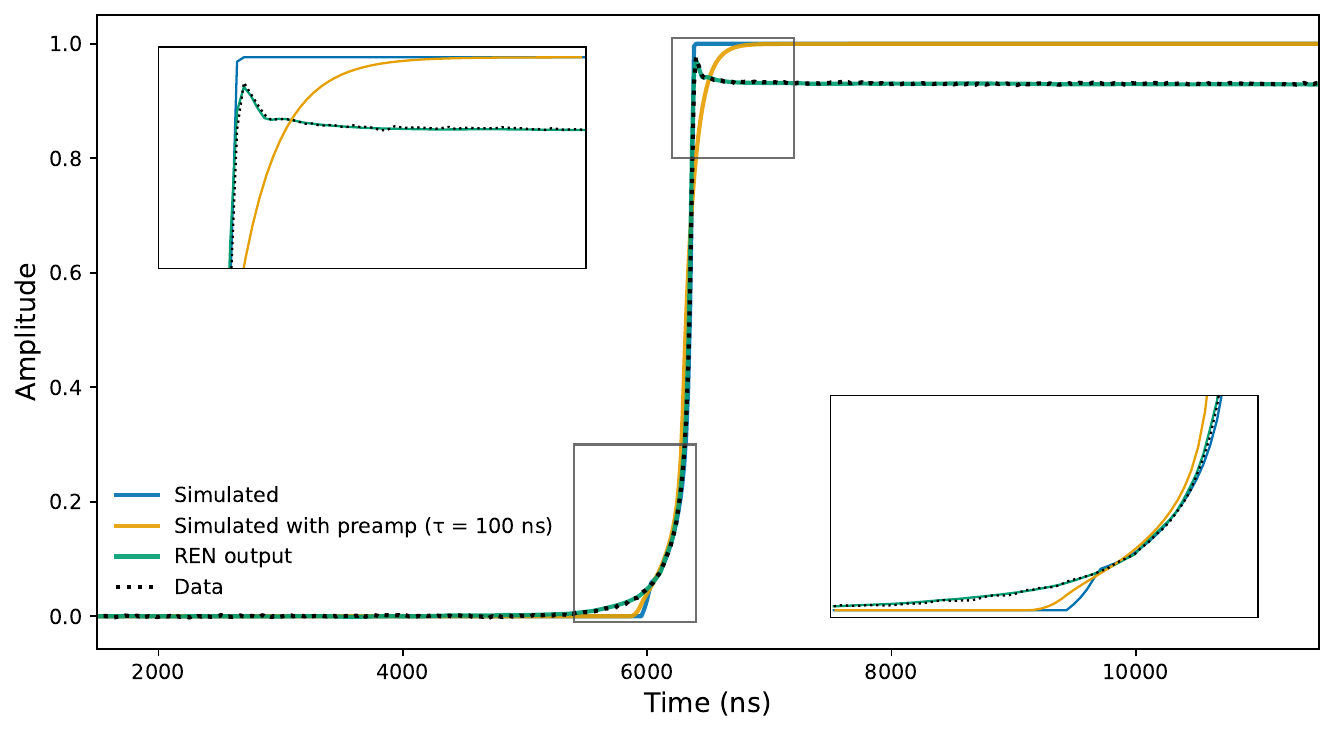}
\caption{Comparison of a single-site event pulse using \siggen\ with no applied electronics effects, \siggen\ with the preamplifier time constant applied (using $\tau = 100$\,ns), the REN-translated version of the \siggen\ pulse with no applied electronics effects, and a comparable data pulse. The REN method improves on the match to data, relative to the standard electronics modeling method.}
\label{fig:pulse_comparision}
\end{figure}

Figure~\ref{fig:pulse_comparision} shows the comparison of one single-site REN-translated pulse with the original {\siggen} simulation input, the {\siggen} simulation with the preamplifier electronics effect enabled, and a comparable data pulse. The two {\siggen} simulated pulses are generated from the same energy deposition information, and the REN-translated pulse is produced from the trained network using the {\siggen}-simulated pulse without the electronics effect enabled as the input. In the absence of labels, there is no way to ensure that the data pulse shown is associated with the same energy deposition location and topology as the simulated pulses, but a single-site data pulse with similar drift time to the simulated pulse has been selected for comparison. All pulses are shown without the pole-zero decay of the tail; the decay is not applied in the two {\siggen} output pulses, and is corrected for by fitting a single decay time constant to the tail in the REN-translated and data pulses. For the {\siggen} simulation with electronics enabled, we applied the built-in {\siggen} preamplifier time constant to the pulse, tuning the time constant to match the mean simulated single-site current amplitude to the mean found in data ($\tau_{rise} = 100$\,ns).

From figure~\ref{fig:pulse_comparision} is is clear that the REN translation provides an improved match relative to the standard simulation approach. Higher-order effects like the pulse overshoot and the smoothing of the initial pulse rise are correctly reproduced by the REN method, but absent from the \siggen-produced pulses using the available built-in electronics modeling. 
 
\subsection{Validation of Key Pulse Parameters}

The training and evaluation sets are unpaired and there is no known one-to-one correspondence between any detector pulse and any simulated pulse. Therefore, a direct event-level comparison is not possible. We assess agreement statistically at the distribution level for topology–informative parameters: drift time $T_{\mathrm{drift}}$, maximum current amplitude $I_{\max}$, and tail decay constant $\tau$.
 
We use the Intersection over Union (IoU) metric, also known as the Jaccard index, to measure the overlap between distributions $A$ and $B$ \cite{murphy1996finley, jaccard_index}. We first calculate the histogram of the distribution being compared using the same binning. The \textit{intersection} is $\sum_{i} \min(\text{bin}_i^A, \text{bin}_i^B)$, the \textit{union} is $\sum_{i} \max(\text{bin}_i^A, \text{bin}_i^B)$, and
 
\begin{equation}
  \text{IoU} \;=\; \frac{\sum_{i} \min(\text{bin}_i^A, \text{bin}_i^B)}{\sum_{i} \max(\text{bin}_i^A, \text{bin}_i^B)}.    
\end{equation}
We express the result in percentages. An IoU of $100\%$ indicates perfect agreement, and $0\%$ indicates no agreement between the histograms.

\subsubsection{Drift Time Distribution}
The electric field of the point-contact style HPGe detector allows events at different locations to have unique pulse shapes, allowing reconstruction of the event topology. Overall, data pulses have a longer drift time than simulations because of the integration of the preamplifier. The metric $tp_{1}$ is defined as the time the pulse reached $1\%$ of its maximum value, and $tp_{100}$ represents the time the pulse reached its maximum value. The time points are found by first finding the maximum time point and then searching backward to find the point when the pulse first crosses that amplitude. The drift time $T_{drift}$ of the pulse is the time between the two points. 

\begin{figure}
\centering
\includegraphics[width=0.49\linewidth,trim={0.8cm 0pc 1.8cm 0pc},clip]{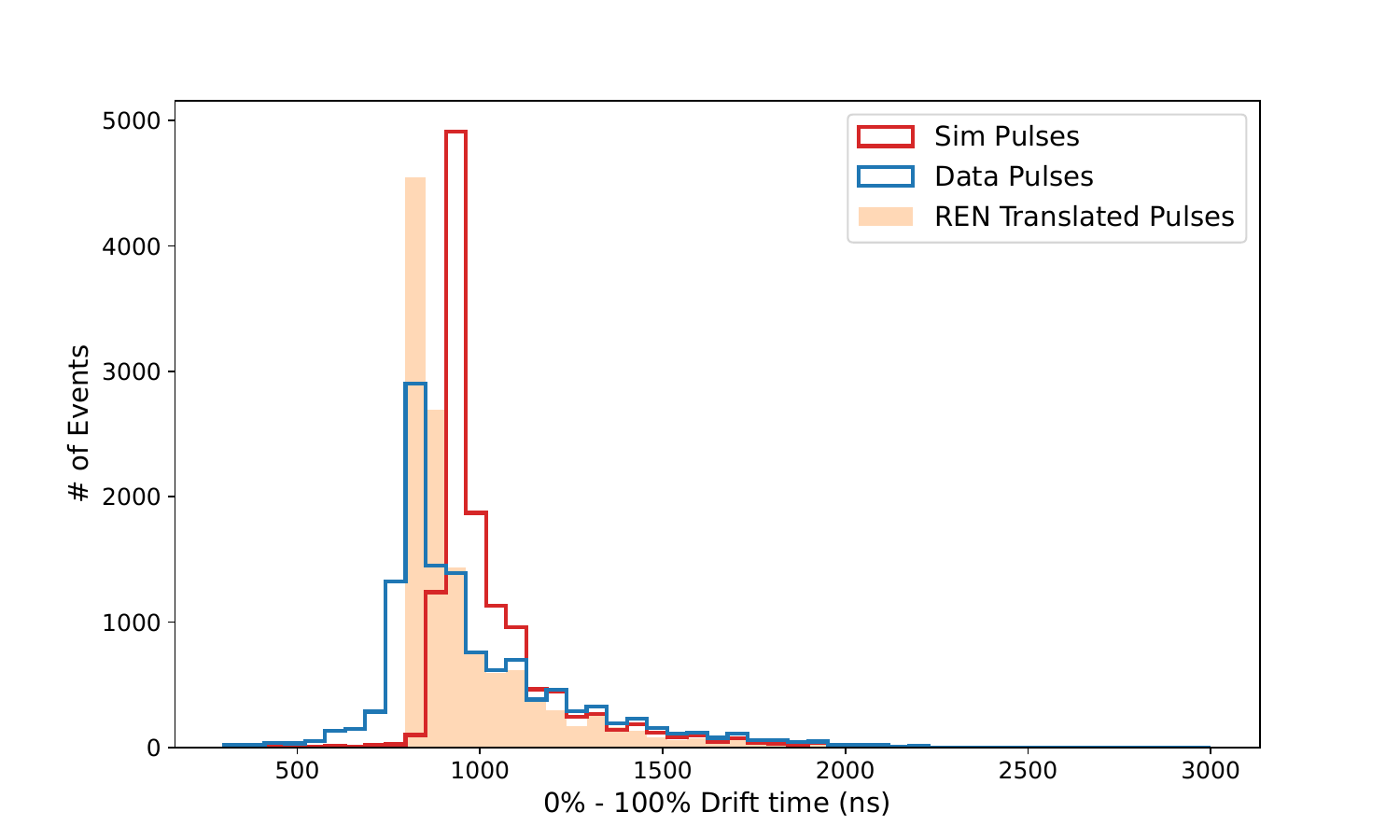}
\includegraphics[width=0.49\linewidth,trim={0.8cm 0pc 1.8cm 0pc},clip]{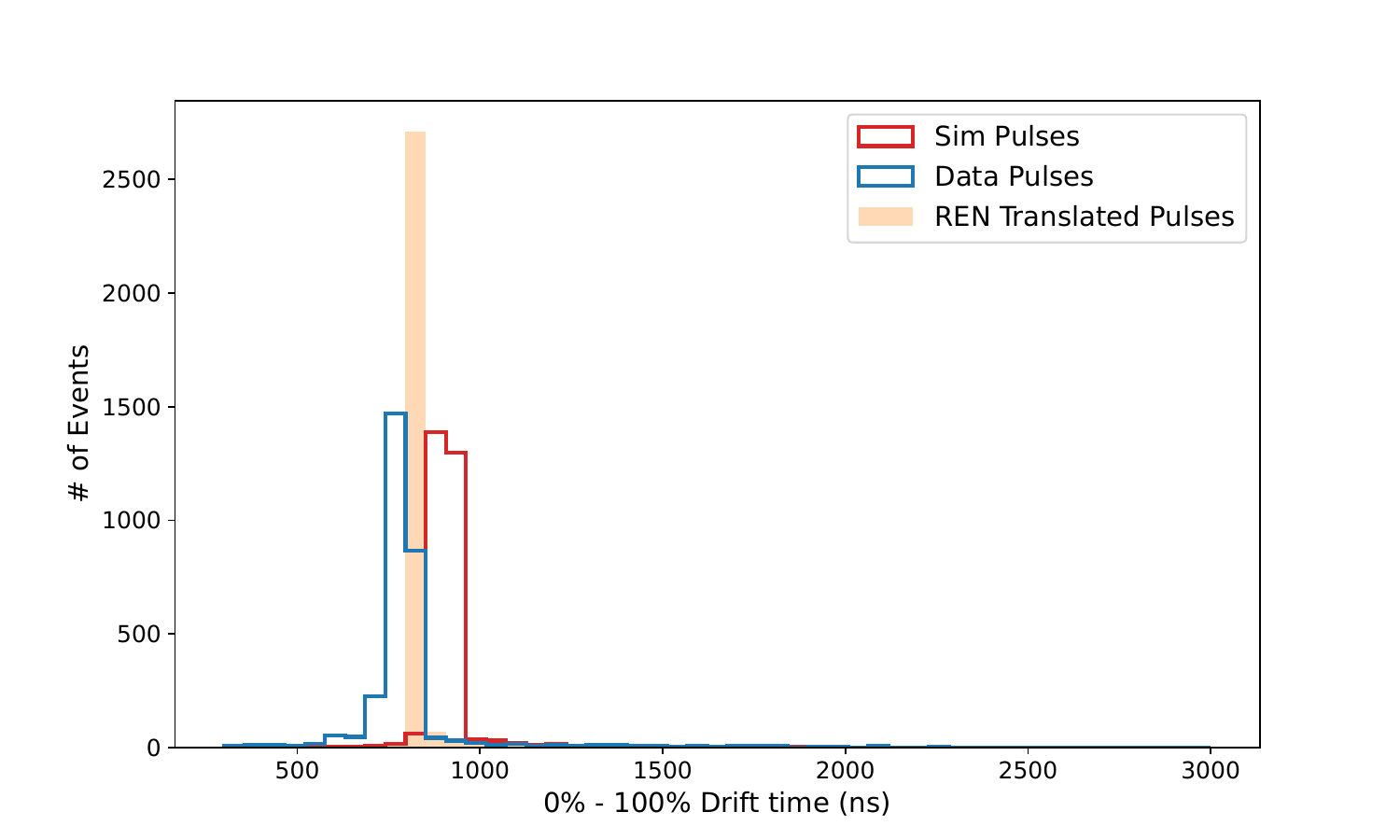}
\caption{The distribution of the $T_{drift}$ on SEP (left) and DEP (right) datasets.}
\label{fig:drift_times_sep}
\end{figure}

Figure \ref{fig:drift_times_sep} illustrates the distribution of $T_{drift}$ in the SEP and DEP validation datasets. REN successfully learns to slow the drift time of the simulated pulses to match the distribution of the data.  The SEP $T_{drift}$ IoU increases to $62.3\%$ from $39.5\%$, a factor of 1.6 improvement. The DEP IoU increases to $22.5\%$ from $5.4\%$, a factor of 4 improvement.
 
\subsubsection{Current Amplitude}
The maximum current amplitude $I_{max}$ is determined by differentiating the pulse and identifying the maximum value of its derivative. For a given event energy, single-site events produce a localized energy deposition, resulting in a sharper and faster increase in $I_{max}$. In contrast, multi-site events, characterized by energy deposition at multiple locations within the detector, yield the same current spread over multiple peaks, decreasing the magnitude of the maximum current value. This distinction makes $I_{max}$ highly effective in differentiating single-site from multi-site events, thereby establishing it as a crucial parameter in pulse shape simulations \cite{AvsE}.

\begin{figure}[!htbp]
\centering
\includegraphics[width=0.49\linewidth,trim={1.1cm 0pc 1.7cm 0pc},clip]{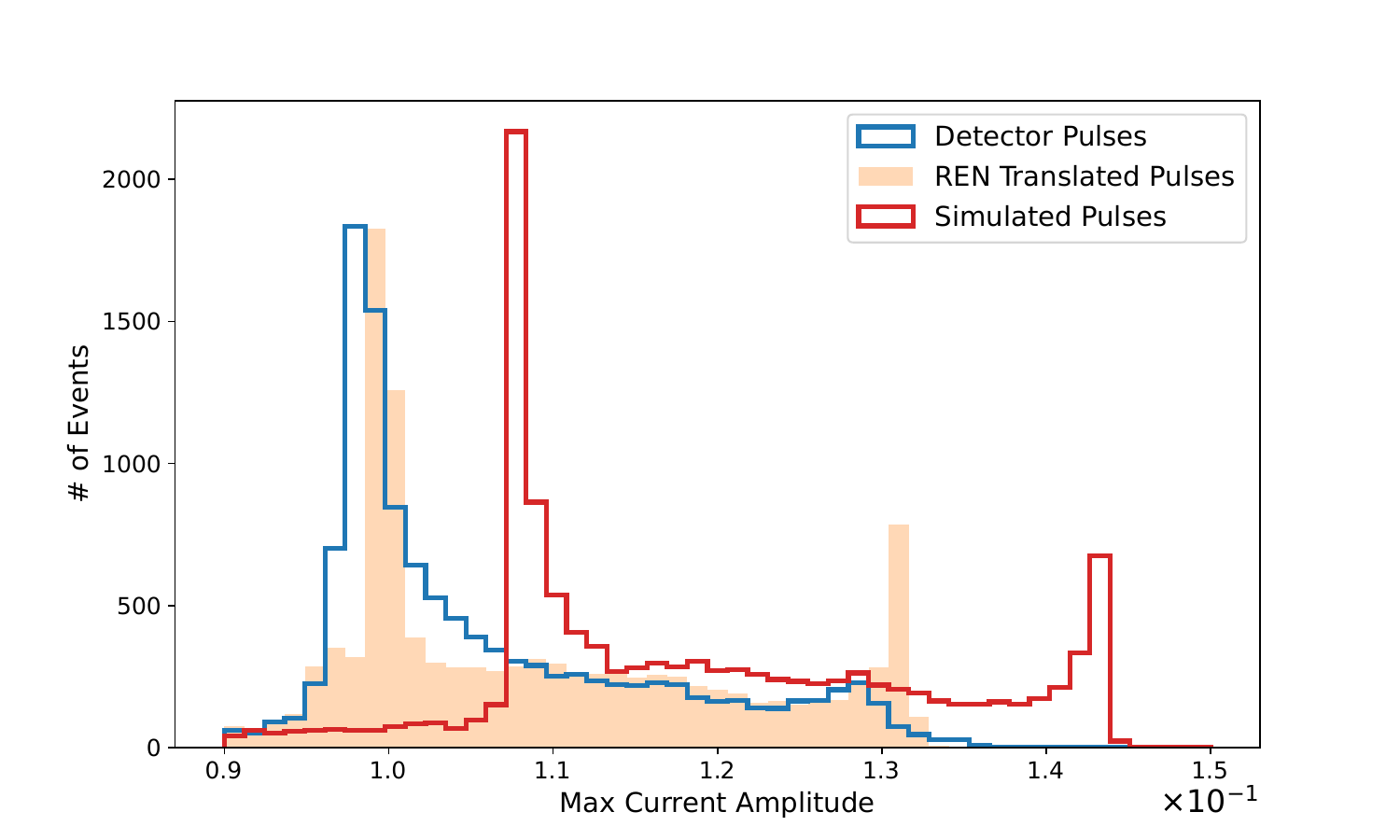}
\includegraphics[width=0.49\linewidth,trim={1.1cm 0pc 1.7cm 0pc},clip]{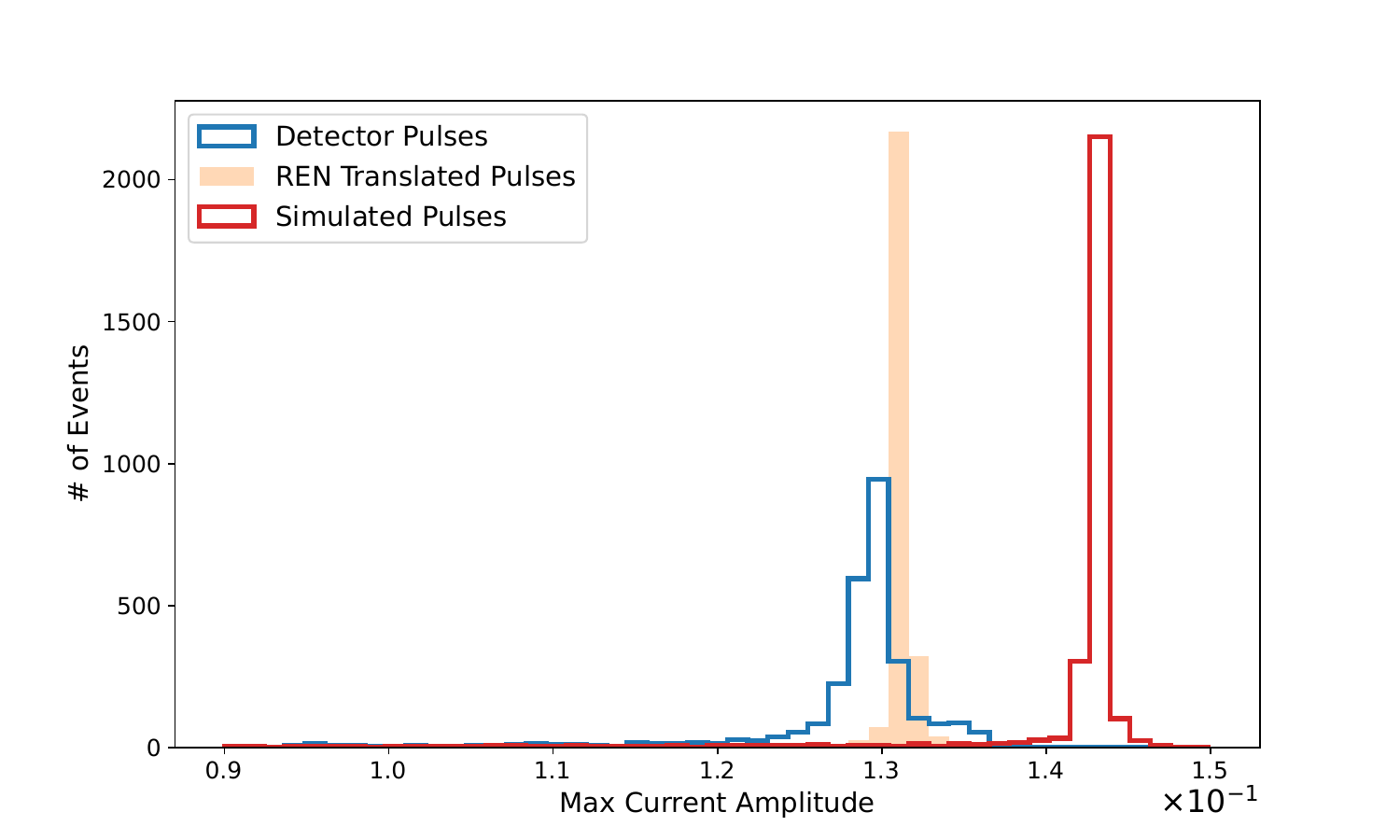}
\caption{ Distribution of maximum current amplitude ($I_{max}$) on SEP (left) and DEP (right) validation datasets.}
\label{ch8_fig_current_amp_sep}
\end{figure}

Figure \ref{ch8_fig_current_amp_sep} illustrates the distribution of $I_{max}$ for both validation datasets. The SEP dataset distributions have two peaks, the higher of which corresponds to single-site events. The peak at a lower value of I$_{max}$ corresponds to multi-site events where one gamma ray from the positron annihilation, with 511 keV, is fully absorbed at a second site. The DEP distribution has only one peak, corresponding to single-site events. REN learns to correctly slow the current amplitude of the simulated pulses to align with the data, while accurately preserving the difference between single-site and multi-site events. The SEP $I_{max}$ increases to the IoU of $63.7\%$ from $27.5\%$, a factor of 2.3 improvement. The DEP IoU increases to $15.5\%$ from $4.2\%$, a factor of 3.7 improvement over simulated pulses with no applied translation.
  
\subsubsection{Tail Slope}
The strength of the RC decay can be measured by the mean tail slope parameter $c_{tail}$. Since the RC decay is an exponential decay, the decay constant $\tau$ was calculated by a linear fit of the logarithm of 300 samples of the pulse. The simulation pulses do not have an RC decay and thus have an infinitely long $\tau$. The data pulses were found to have a mean $\tau$ of $54.60 \mu s$, while the REN translated pulses had a mean $\tau$ of $53.75 \mu s$, deviating from the measured value by only $1.6\%$.

\subsection{Discussion}
Our use case of the Cycle-GAN technique allows for quantitative studies of its performance using well-studied and physically meaningful parameters. In addition to these studies, we can also conduct a qualitative evaluation. Figures~\ref{fig:pulse_examples} and \ref{fig:pulse_comparision} show that the data-like pulse shapes resulting from applying {\cpunet} include many of the more-subtle features seen in the data, such as overshoot at the end of each rising edge and the impact of multiple RC decay constants. This suggests that as novel PSD parameters and techniques are developed, simulated pulses with {\cpunet} translation applied to them will serve as an accurate model for true data pulses. 

{\cpunet} seeks to determine the transformation $\Lambda$ that will take each simulated pulse to the equivalent realistic detector pulse; its structure is, by design, blind to ensemble-level features like the width of peaks in the $T_{drift}$ and $I_{max}$ distributions, which emerge from variations in these parameters among pulses rather than an overall effect applied to all pulses. Effects like the overall too-narrow peaks observed in the simulations used for this work have been connected to unmodeled physics effects in the \siggen\ software package. Tests employing a more-realistic charge cloud model, for example, show broader peaks in these distributions, improving agreement with the data. These ensemble-level discrepancies are preserved by {\cpunet}, so the overall network performance on the IoU metrics has the potential to improve with the use of more realistic pulse simulations.  

\subsection{Uncertainty Evaluation}\label{App:uncertainty}
\begin{figure}[!htbp]
\centering
\includegraphics[width=0.495\linewidth,trim={0.1cm 0.2cm 0.2cm 0.8cm},clip]{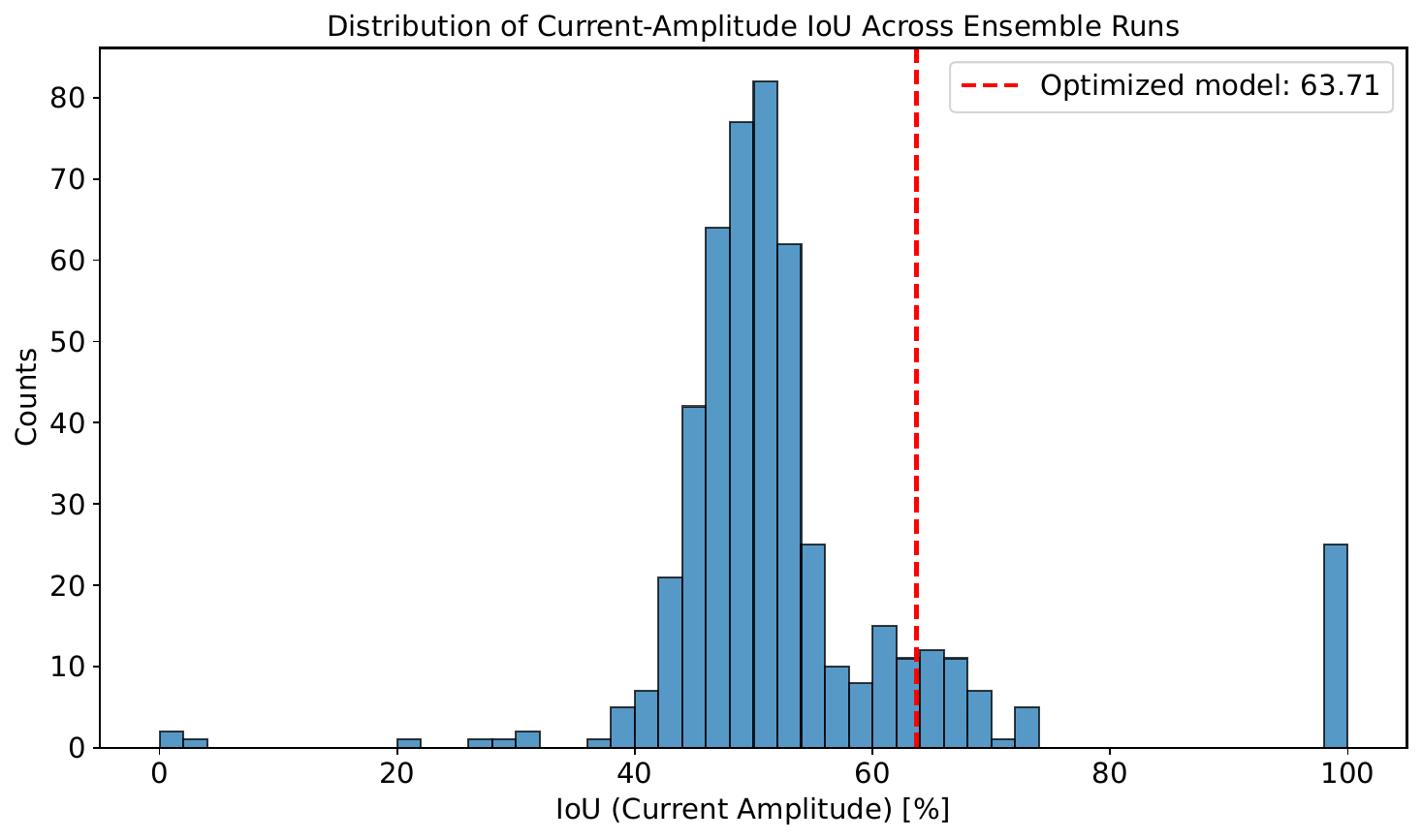}
\includegraphics[width=0.485\linewidth,trim={0.1cm 0.2 0.2cm 0.8cm},clip]{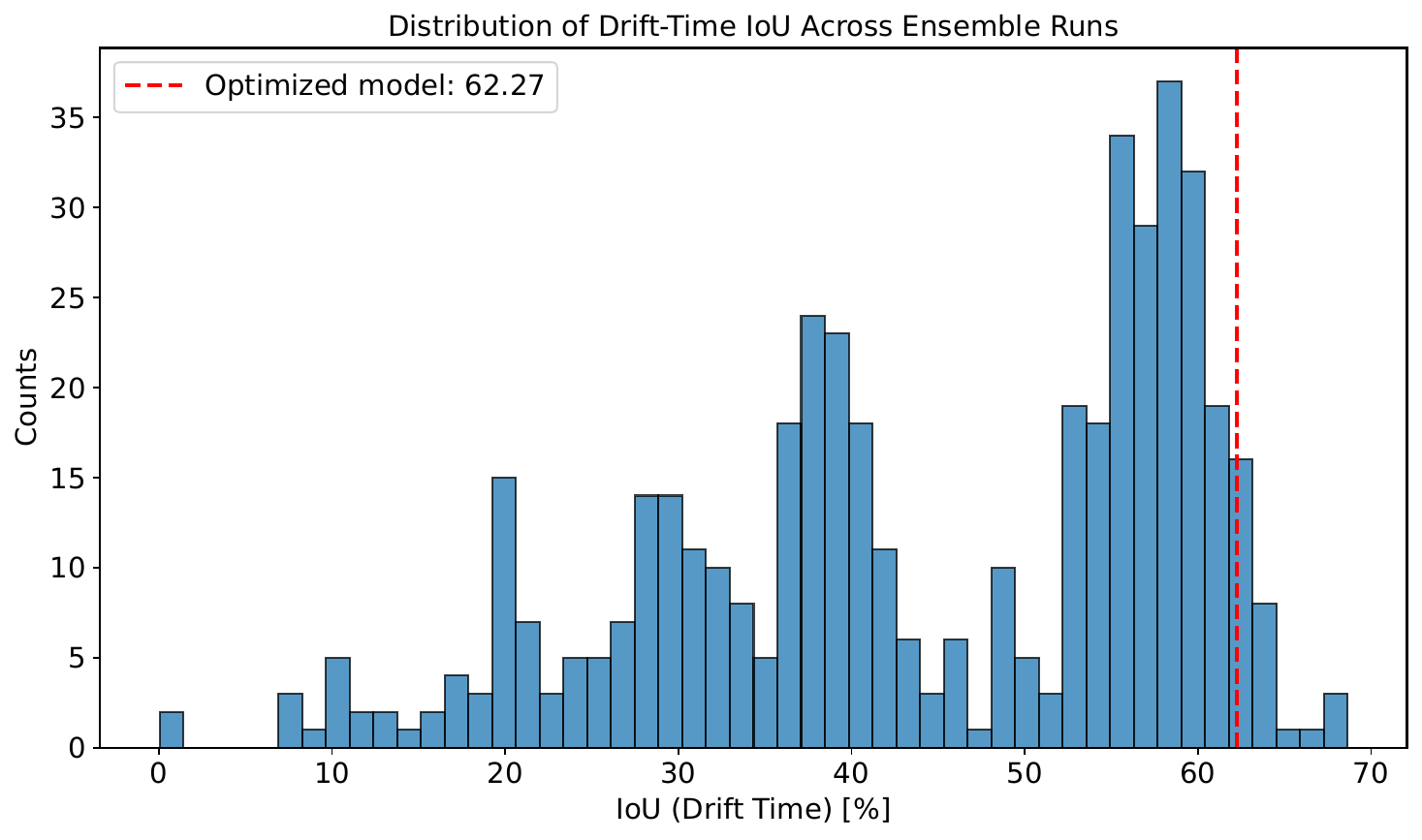}
    \caption{IoU distribution of 500 CPU-Net trials, varying the weight initialization, over maximal current amplitude $I_{max}$~(left) and drift time $T_{drift}$~(right). The distributions quantify the spread in performance arising from stochastic optimization and initialization, providing an estimate of the uncertainty of CPU-Net performance for fixed hyperparameters. The multi-peaked structure reflects that the adversarial training tends to converge into a small number of distinct modes in local minima in the parameter space, rather than a single unique solution. The low IoU tail corresponds to runs in which the GAN training becomes unstable, for instance when the generator or discriminator overpowers its counterpart, which lead to poorer alignment between translated and detector distributions. The dashed red lines indicate the IoU values of the optimized model used in the main analysis, which was obtained during hyperparameter tuning with a fixed random seed.}
    \label{fig:uncert}
\end{figure}

To measure the uncertainty associated with the network initialization process, we trained 500 trials of CPU-Net using different random seeds. The IoU over $I_{max}$ and $T_{drift}$ are then calculated for each trial over the same SEP validation dataset of 12,000 pulses. Figure~\ref{fig:uncert} shows the distribution of IoU values for the current amplitude and drift time distributions. The optimized network results, also shown in this figure, are biased towards higher values of the IoU metrics because hyperparameter tuning was performed using a fixed seed value. 

The multi-peaked structure in these distributions appears to be a consequence of stochastic training rather than a physical effect. Each point in those histograms is an independent training run with a different random seed, and the adversarial training appears to fall into a few distinct convergence modes. These modes can be interpreted as different local minima in the $\sim 14.2$M parameter space of the full model (two generators and two discriminators).

Since many of the hyperparameters are tuned to encourage network convergence when starting from the initial weights associated with that specific seed, varying the seed value can lead to failed GAN training, where one component of the adversarial network overpowers another. This indicates that beginning from pre-trained network weights may be a helpful strategy in scaling CPU-Net so it can be applied to multiple detectors and varying run conditions. 

%trim= left bottom right top

\section{Conclusion}\label{sec:conclusion}
\subsection{Summary of Results}
Pulse shape simulation is pivotal to improving analysis techniques in experiments that rely on HPGe detectors. A major challenge in these simulations is to correctly model the effects of the readout electronics. We developed a CycleGAN network that learns to translate simulated pulse shapes into data-like signals without explicitly modeling the electronics. The model combines a positional U-Net generator with an RNN-based discriminator, and is trained on unpaired data sets of simulated and data pulses that are collected during routine calibration measurements. {\cpunet} successfully translates simulated pulses to output pulses that closely resemble actual detector pulses by applying corrections using a neural network. We show that {\cpunet} preserves the necessary physics in the pulses and correctly reproduces critical pulse shape parameters.

Imperfect simulations are a common problem facing physics experiments; one potential source of mismatch is inaccurate modeling of the electronics response inherent in signal readout systems. The architecture we have developed is flexible enough to be used for any time-series data where first-principles simulations exist, but do not perfectly match true data. The same method can be adapted to time-series image data by replacing the RNN discriminators with a Convolutional Neural Network (CNN)-based method, and adapting the U-Net architecture as needed. By using a data-driven electronics emulation method, {\cpunet} resolves a longstanding obstacle in PSS: the complexity of replicating the detector’s electronic response.

\subsection{Future Work}
Although our main focus in this work was to translate simulations to resemble measured data, {\cpunet}’s bidirectional nature also offers the potential for denoising real detector pulses. By transforming noisy detector signals into cleaner, simulation-like pulses, {\cpunet} can help isolate and identify subtle features in the event topology. This capability could ultimately improve spatial reconstruction and improve our understanding of particle interactions within the detector.

 HPGe detector measurements using the Compton scanning technique provide position-tagged, paired waveforms using a Compton-camera setup \cite{Abt_2022odr}. Using data from such a system would enable direct quantitative per-pulse evaluation of {\cpunet}. We leave this dedicated study for future work. Future studies will also study how {\cpunet} generalizes to other detectors, geometries, and changing experimental conditions.

\section{Acknowledgment}\label{sec:ack}
This work was supported by the U.S. Department of Energy, Office of Science, Office of Nuclear Physics, under award numbers DE-SC0022339, DE-FG02-97ER41041, and DE-FG02-97ER41033, and by the National Science Foundation under grant PHY-1812374. 

This material is based upon work supported by the U.S. Department of Energy, Office of Science, Office of Workforce Development for Teachers and Scientists, Office of Science Graduate Student Research (SCGSR) program. The SCGSR program is administered by the Oak Ridge Institute for Science and Education for the DOE under contract number DE‐SC0014664. 

We gratefully acknowledge contributions from \textsc{Majorana Demonstrator} and the LEGEND experiments, and we thank our collaborators for their input. Part of this research was carried out at the Aspen Center for Physics, which is supported by NSF grant PHY-1607611. We also thank Dr. Michelle Kuchera for insightful discussions.

\bibliographystyle{unsrt}
\bibliography{cpu_net}

\end{document}